\RequirePackage{rotating}
\documentclass[useAMS]{mn2e}
\usepackage{graphicx}
\usepackage{longtable}
\usepackage{rotate}
\usepackage{rotating}
\usepackage{mathtools}
\usepackage{pdflscape}
\usepackage{scalerel}
\usepackage{times}
\usepackage{verbatim}
\newif\ifAMStwofonts

\def\gs{\mathrel{\hbox{\rlap{\hbox{\lower4pt\hbox{$\sim$}}}\hbox{$>$}}}}
\def\ls{\mathrel{\hbox{\rlap{\hbox{\lower4pt\hbox{$\sim$}}}\hbox{$<$}}}}

\def\suzaku{{\it Suzaku}}
\def\nustar{{\it NuSTAR}}

\def\hst{{\it HST}}

\def\swift{{\it Swift}}
\def\xmm{{\it XMM-Newton}}

\def\et{{et al.\ }}

\def\mrk335{{Mrk~335}}
\def\iras13{{IRAS~13224--3809}}
\def\1h07{{1H~0707--495}}
\def\izw1{{I~Zw~1}}
\def\rg{{\thinspace r_{\rm g}}}

\def\Cdof{{C/{\rm dof}}}

\def\redchi{{\chi^2_\nu}}

\def\feka{{Fe~K$\alpha$}}

\def\nh{{N_{\rm H}}}

\def\deg{^{\circ}}

\def\cm{{\rm\thinspace cm}}
\def\erg{{\rm\thinspace erg}}
\def\eV{{\rm\thinspace eV}}

\def\ga{{\rm\thinspace gauss}}

\def\keV{{\rm\thinspace keV}}
\def\km{{\rm\thinspace km}}

\def\Msun{\hbox{$\rm\thinspace M_{\odot}$}}

\def\s{{\rm\thinspace s}}
\def\ks{{\rm\thinspace ks}}
\def\ps{{\rm\thinspace s^{-1}}}

\def\cmps{\hbox{$\cm\s^{-1}\,$}}

\def\ergpscmps{\hbox{$\erg\cm^{-2}\s^{-1}\,$}}

\def\kmps{\hbox{$\km\ps\,$}}

\def\pscm{\hbox{$\cm^{-2}\,$}}

\title[Winds and jets in \mrk335]
      {
Evidence for an emerging disc wind and collimated outflow during an X-ray flare in the narrow-line Seyfert 1 galaxy \mrk335     }

\author[L. C. Gallo \et]
       {
       L. C. Gallo,$^1$ 
       A. G. Gonzalez,$^1$ 
       S. G. H. Waddell,$^1$
       H. J. S. Ehler,$^1$
       D. R. Wilkins,$^2$
\newauthor
	A. L. Longinotti,$^3$
	D. Grupe,$^4$
	S. Komossa,$^5$
       G. A. Kriss,$^6$
       C. Pinto,$^7$
       S. Tripathi,$^1$
\newauthor
        A. C. Fabian,$^7$
       Y. Krongold,$^8$
       S. Mathur,$^9$
       M. L. Parker,$^{10}$
and       A. Pradhan,$^9$  
        \\ 
$^{1}$ Department of Astronomy and Physics, Saint Mary's University, 923 Robie Street, Halifax, NS, B3H 3C3, Canada \\
$^{2}$ Kavli Institute for Particle Astrophysics and Cosmology, Stanford University, Stanford, CA, 94305, U.S.A. \\
$^{3}$ Instituto Nacional de Astrof\'isica, \'Optica y Electr\'onica, Luis E. Erro 1, Tonantzintla, Puebla, M\'exico \\
$^{4}$ Space Science Center, Morehead State University, 235 Martindale Drive, Morehead, KY 40351, USA\\
$^{5}$ Max-Planck-Institut f{\"u}r Radioastronomie, Auf dem H\"ugel 69, 53121 Bonn, Germany \\
$^{6}$ Space Telescope Science Institute, 3700 San Martin Drive, Baltimore, MD 21218, USA  \\
$^{7}$ Institute of Astronomy, University of Cambridge, Madingley Road, Cambridge CB3 0HA\\
$^{8}$ Instituto de Astronomia Universidad Nacional Autonoma de Mexico, Mexico City, Mexico\\
$^{9}$ Department of Astronomy, The Ohio State University, 140 West 18th Avenue, Columbus, OH 43210, USA\\
$^{10}$ European Space Agency (ESA), European Space Astronomy Centre (ESAC), Villanueva de la Ca\~n–ada, E-28691 Madrid, Spain \\
}
\date{Accepted. Received. }
\pagerange{\pageref{firstpage}--\pageref{lastpage}}
\pubyear{2010}
\begin{document}
\maketitle
\label{firstpage}

\begin{abstract}
A triggered $140\ks$ \xmm\ observation of the narrow-line Seyfert 1 (NLS1) \mrk335\ in December 2015 caught the active galaxy at its lowest X-ray flux since 2007.Ê The NLS1 is relatively quiescent for the first $\sim120\ks$ of the observation before it flares in brightness by a factor of about five in the last $20\ks$.Ê Although only part of the flare is captured before the observation is terminated, the data reveal significant differences between the flare and quiescent phases.Ê During the low-flux state, \mrk335\ demonstrates a reflection-dominated spectrum that results from a compact corona around a Kerr black hole.Ê In addition to the rapid brightening, the flare is further described by spectral softening and a falling reflection fraction that are consistent with previous observations advocating at least part of the corona in \mrk335\ could be the base of an aborted jet. Ê
The spectrum during the flaring interval reveals several residuals between the $2-3\sigma$ level that could be attributed to absorption lines from a highly ionised plasma that is moving outward at $v\sim0.12c$.Ê It could be that the increased luminosity during the flare enhances the radiation pressure sufficiently to launch a possible wind.Ê If the wind is indeed responding to the change in corona luminosity then it must be located within $\sim80\rg$.Ê The escape velocity at this distance is comparable to the estimated wind velocity.Ê If confirmed, this is the first example of a radio-quiet AGN exhibiting behaviour consistent with both diffuse and collimated outflow.

\end{abstract}



\section{Introduction}
\label{sect:intro}

While the details are uncertain, it is true that the energy expelled from the central engine of active galactic nuclei (AGN) can play a significant role in galaxy evolution (e.g. Hopkins 2012; Fabian 2012).  The ejection mechanism is clearly visible in many AGN, whether it be the collimated, relativistic jets in some radio-loud objects (e.g. Bridle \& Perley 1984) or diffuse accretion disc winds seen in the UV spectral features of some radio-quiet AGN (e.g. Murray \et 1995).  The more powerful relativistic jets may draw energy from a rapidly rotating black hole at the centre (e.g. Blandford \& Znajek 1977), whereas the disc winds may be launched from a larger distance by radiation or magnetic pressure (e.g. Blandford \& Payne 1982; Murray \et 1995; Proga \& Kallman 2004).  Although uncommon, there are examples of jets and winds co-existing in radio-loud AGN (e.g. Giroletti \et 2017; Tombesi \et 2010, 2012, 2014).
  
It has become clear that even low-luminosity, radio-quiet AGN are capable of launching jets (e.g. Giroletti \& Panessa 2009; Foschini \et 2015; L\"ahteenm\"aki \et 2018) and evidence for winds are seen in the X-ray spectra of several Seyferts (e.g. Krongold \et 2003; Tombesi \et 2010; Longinotti \et 2015).  Both processes are variable and may be working intermittently in some objects (e.g. Matzeu \et 2016; Parker \et 2017; Gonzalez \et 2017a).

In recent years, a number of AGN, specifically narrow-line Seyfert 1 galaxies (NLS1s; e.g. Gallo 2018; Komossa 2018), exhibit behaviour that could be reconciled if some significant part of the corona were attributed to the base of a jet (e.g. Gallo \et 2007, 2013; Wilkins \et 2015, 2017).   Ghisellini \et (2003) propose that radio-quiet Seyferts may be hosting aborted jets that work intermittently, and whose ejecta do not reach the escape velocity of the system, but rather fall back colliding with other ejected blobs.

This description of the X-ray emitting region has been especially appropriate for the NLS1 \mrk335\ ($z=0.025$).  \mrk335\ has been extensively studied since falling to an X-ray dim state in 2007 (e.g. Grupe \et 2007, 2008).  Over the past twelve years, \mrk335\ has exhibited persistent flickering and occasional high-amplitude flares (e.g. Grupe \et 2012; Wilkins \et 2015; Gallo \et 2018).  While the low flux state appears consistent with a compact corona (e.g. Grupe \et 2007; Gallo \et 2013, 2015; Parker \et 2014; Wilkins \& Gallo 2015), the flares have been attributed to a corona that may be attempting to launch material as in the aborted jet scenario (e.g. Gallo \et 2013, 2015; Wilkins \et 2015).  During these flaring events, X-ray spectral modelling suggests the compact corona is beaming emission away from the disc.  

In December 2015, \mrk335\ dropped to its lowest X-ray flux ever recorded triggering a \xmm\ and \hst\ Target-of-Opportunity observation (PI: Longinotti) to study the UV and warm absorbers in the X-ray low state (Longinotti \et in prep). 
Here, the broadband X-ray data are examined with the blurred reflection model, which show evidence for the simultaneous onset of a collimated outflow and an accretion disc wind in the NLS1.


\section{Observation and data reduction}
\label{sect:data}

\mrk335\ was observed for $\sim140\ks$ with \xmm\ (Jansen \et 2001) starting 30 December, 2015 as part of a coordinated campaign with the Hubble Space Telescope (\hst) to study the ionised absorbers and emitters in the UV and X-ray low-flux state.  The observation was triggered by \swift\ monitoring (Grupe \et 2015) and the \xmm-\hst\ analysis is reported by Longinotti \et (in prep).

The \xmm\ data from the European Photon Imaging Camera (EPIC) are presented here.  During the observations the EPIC detectors (Str\"uder \et 2001; Turner \et 2001) were operated in full-window mode and with the thin filter in place.  The \xmm\ Observation Data Files (ODFs) 
were processed to produce calibrated event lists using the \xmm\ Science Analysis System ({\sc SAS v16.0.0}).  Light curves were extracted from these event lists to search for periods of high background flaring.  Such periods were neglected resulting in a good-time exposure of $\sim117\ks$.

Spectra were extracted from a $35$ arcsec circular region centred on the source.  
The background photons were extracted from an off-source region on the same CCD.
Pile-up was negligible during the observations.
Single and double events were selected for the pn detector, and
single-quadruple events were selected for the Metal Oxided Semi-conductor (MOS) detectors.
EPIC response matrices were generated using the {\sc SAS}
tasks {\sc ARFGEN} and {\sc RMFGEN}.  
The MOS and pn data were compared for consistency and determined to
be in agreement within known uncertainties.

The spectra were optimally binned following Kaastra \& Bleeker (2016).
Spectral fitting was performed using {\sc XSPEC v12.9.1}
(Arnaud 1996) and fit quality is tested using the $C$-statistic (Cash 1979).
All parameters are reported in the rest frame of the source unless specified
otherwise, but figures remain in the observed frame.
The quoted errors on the model parameters correspond to a 90\% confidence
level for one interesting parameter.
A value for the Galactic column density toward \mrk335\ of
$3.56 \times 10^{20}\pscm$ (Kalberla \et 2005) is adopted in all of the
spectral fits and
abundances are from Wilms \et (2000).

\section{Spectral variability}
\label{sect:spec}

\subsection{Flux resolved spectroscopy}

\begin{figure}
\rotatebox{0}
{\scalebox{0.35}{\includegraphics[trim= 1.2cm 1.4cm 1.5cm 3cm, clip=true]{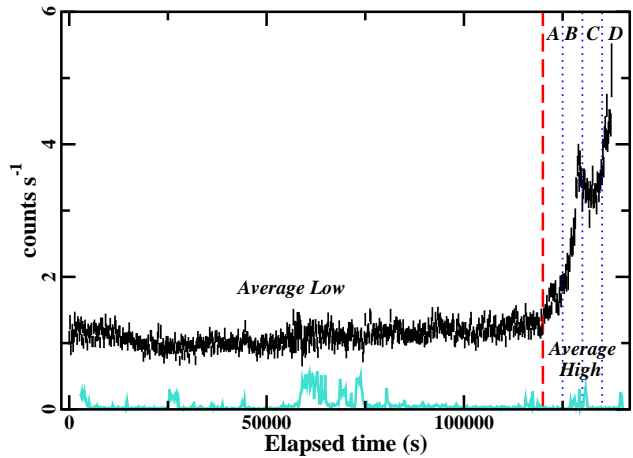}}}
\caption{
The $0.2-10\keV$ EPIC-pn count rate of \mrk335\ during the 2015 low-flux state.  Six spectra are created based on the source flux.  The vertical dashed line (red) delineates the time intervals from which the average low-flux ($<120\ks$) and flare (i.e. high-flux) ($>120\ks$) spectra are created.  Data during the X-ray flare are divided into four segments ($\sim5\ks$ each) of increasing flux from $A - D$ (shown by the dotted blue lines).  The background level is shown for comparison (lower curve, turquoise).
}
\label{fig:lc}
\end{figure}
\begin{figure}
\rotatebox{0}
{\scalebox{0.35}{\includegraphics[trim= 0.9cm 0.8cm 1.5cm 2cm, clip=true]{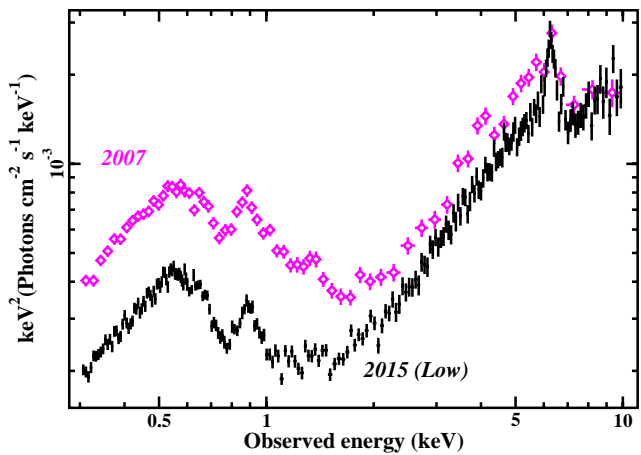}}}
\caption{
The EPIC-pn spectrum of \mrk335\ during the 2007 low flux observation (magenta diamonds) compared to the low-flux interval ($\ls120\ks$ in Fig.~\ref{fig:lc}) in 2015 (black crosses).
}
\label{fig:lowspec}
\end{figure}

The $0.2-10\keV$ light curve of \mrk335\ shows the NLS1 exhibiting quiescent behaviour in the first $\sim120\ks$ of the observation before the onset of a large amplitude flare in the last $\sim20\ks$ (Fig.~\ref{fig:lc}).  During the flaring period, the count rate increases by at least a factor of five before the observation ends.  During the average low-flux interval, \mrk335\ is at its lowest flux ever observed with \xmm.  For comparison, the spectrum during the first $120\ks$ is compared to the previous low-flux state captured in 2007 (Fig.~\ref{fig:lowspec}).  The 2015 dim state shows overall lower flux across the band and marked hardening below $\sim2\keV$ compared to the 2007 low state.

The 2015 data are divided to create spectra in six flux intervals as marked in Fig.~\ref{fig:lc}.  There is an average low state constituting data before $\sim120\ks$ in the light curve and a flare spectrum (average high) that includes data after $120\ks$.  Data during the flare are divided into four, $5\ks$ intervals of increasing flux (A, B, C, and D).  Although, spectrum D contains the highest flux data it is also of the shortest exposure (i.e. good-time interval, $\sim2.2\ks$).  The six spectra are fitted simultaneously. 

Only the blurred reflection model is considered in the examination of these data. The exercise of testing various models for \mrk335\ has been carried out in several earlier works (e.g. Gallo \et 2013, 2015; Grupe \et 2008, 2012).  Gallo \et (2015) demonstrate the blurred reflection model is statistically preferred over partial-covering models.  Partial covering is adopted to describe the continuum in these current data in the work of Longinotti \et (in prep).  The testing of other models on this particular data set is the subject of future work.

In the blurred reflection model (e.g. Ross \& Fabian 2005), the primary continuum source is a corona situated in some geometry above the disc and radiating isotropically in the rest-frame.  Some of the primary emission (power law continuum) will illuminate the accretion disc producing back-scattered emission called the reflection spectrum, which in the \xmm\ energy band, is composed most prominently of fluorescent emission lines.  Arising in the rapidly rotating accretion disc close to the supermassive black hole, the reflection spectrum will be blurred by Keplerian and relativistic (special and general) effects.

This scenario is modeled in {\sc xspec} using the models listed in Table~\ref{tab:fits}.  The model resembles that fitted to the spectra from the  \suzaku\ low-flux state analysis (Gallo \et 2015).  Given the additional sensitivity below $\sim1\keV$ provided by \xmm, the model in this current work also includes a warm absorber and collisionally ionised emitter ({\sc mekal} in {\sc xspec}) that are phenomenologically consistent with the absorption and emission features found in the RGS analysis (Longinotti \et in prep).  

Parameters that are not expected to vary over the decade-long time scales since \mrk335\ has been monitored (e.g. black hole spin, inclination, and iron abundance) have been left free to vary rather than constrained to previously measured values.  The measurements in this analysis are consistent with earlier work, but a multi-epoch analysis (e.g. Wilkins \& Gallo 2015; Keek \& Ballantyne 2016) is left for the future.

The variability between the different flux intervals can be described very simply with changes in only three parameters: the power law normalisation, power law photon index ($\Gamma$), and disc ionisation parameter ($\xi=4\pi F/n$ where $n$ is the hydrogen number density and $F$ is the incident flux).  The model fits the data well ($\Cdof=851/606$, Table~\ref{tab:fits}).  The behaviour is very similar to that previously observed from \mrk335\ (Gallo \et 2013) and other NLS1s (e.g. Bonson \et 2018; Chiang \et 2015).  The fact these parameters vary together is expected in the blurred reflection scenario as the power law is illuminating and ionising the inner accretion disc.   Allowing only one component to vary independent of the other (e.g. either the power law parameters alone or the reflection parameters alone) increases the C-statistic significantly ($\Delta C > 169$ for 10 fewer free parameters).

\begin{figure}
\rotatebox{0}
{\scalebox{0.315}{\includegraphics[trim= 0.7cm 1.2cm 2.5cm 3cm, clip=true]{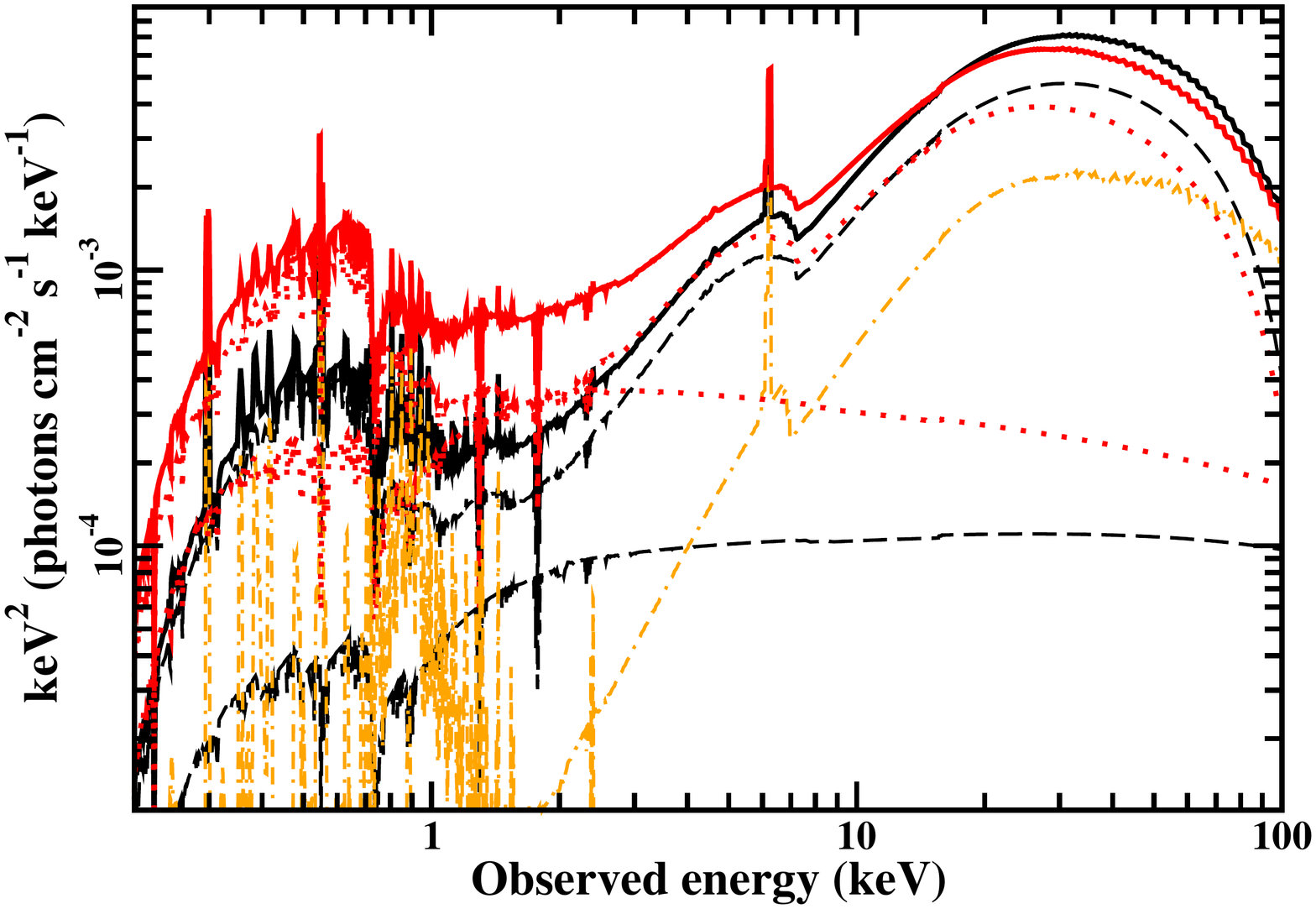}}}
{\scalebox{0.32}{\includegraphics[trim= 1.cm 1.2cm 1.5cm 3cm, clip=true]{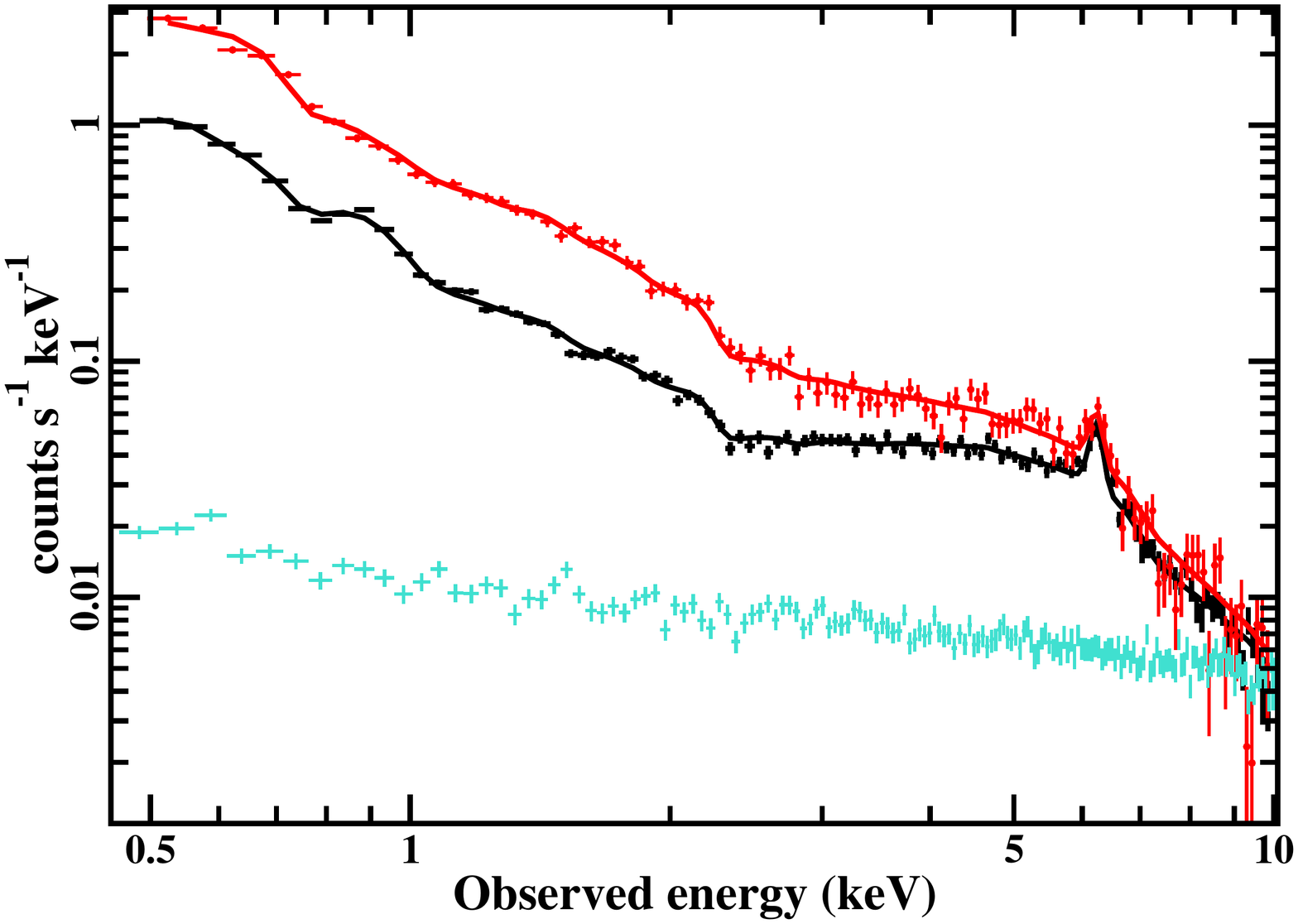}}}
{\scalebox{0.32}{\includegraphics[trim= 1.cm 1.42cm 1.5cm 3cm, clip=true]{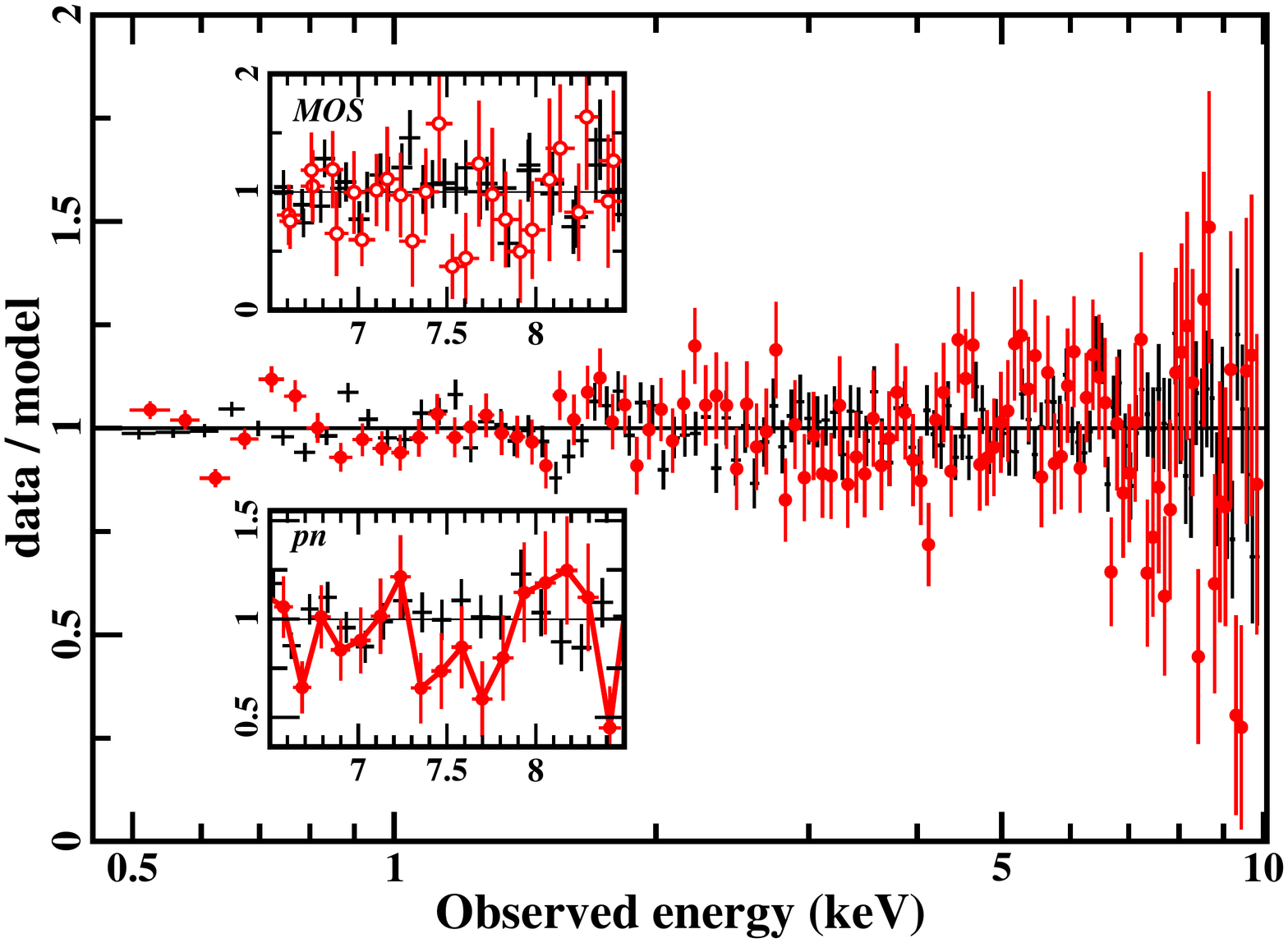}}}
\caption{
The blurred reflection model fitted to the average low- and high-flux states seen in \mrk335\ in 2015.  The only model parameters permitted to vary between the two spectra is the power law photon index ($\Gamma$) and normalisation, and the ionisation parameter ($\xi$) of the reflector.  {\bf Top panel:}  The model components shown between $0.1-100\keV$.  The solid curves are the total model for the low-flux (black) and high-flux (red) spectra.  The low-flux components are shown as (black) dashed curves and the high-flux components are shown as (red) dotted curves.  The (orange) dot-dash curves are the distant reflector and the ionised emitter that are constant between the two spectra.  The power law is steeper during the bright state and the relative flux between the power law and blurred reflector also changes.  {\bf Middle panel:} The observed, background subtracted pn spectra of the low- and high-flux intervals.  Differences are primarily seen below $\sim6\keV$.  The average background is shown for comparison (turquoise) {\bf Lower panel:} The pn ratio (data / model) demonstrating the quality of the model fit.  Both the bright and dim spectra are well-fitted by the blurred reflection model.  The $6.5-8.5\keV$ residuals are enlarged in the lower inset showing a possible absorption feature at $\sim7.5\keV$ that is only in the high-flux spectrum.  The upper inset in the panel shows the residuals in the MOS spectra during the high-flux (red open circles) and low-flux (black crosses), which also show a dip at $E\sim7.5\keV$ during the flare. 
}
\label{fig:meanfit}
\end{figure}

\begin{figure}
\rotatebox{0}
{\scalebox{0.35}{\includegraphics[trim= 1.cm 1.4cm 1.4cm 1cm, clip=true]{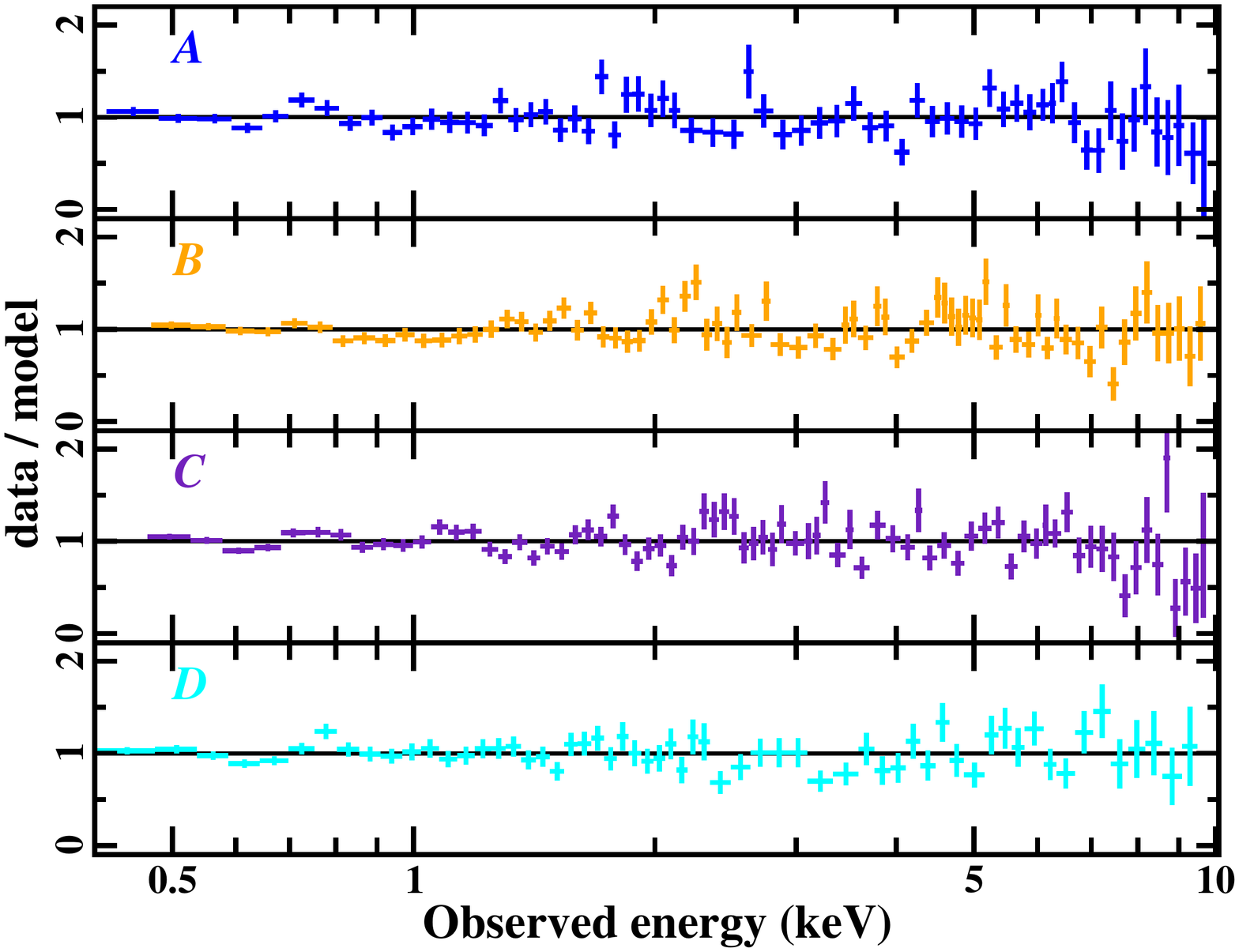}}}
\caption{
The residuals (data/model) from fitting the blurred reflection model to the flux-resolved flare spectra noted in Fig.~\ref{fig:lc}.  The flux increases from $A$ (top panel) to $D$ (lowest panel) over the duration of the flare.  As the flux increases, an increasing trend is seen in the photon index and ionisation parameter.
}
\label{fig:highfit}
\end{figure}

The most prominent residuals that remain are seen between $7-8\keV$ in the average flare spectrum (lower panel, Fig.~\ref{fig:meanfit}).  The residuals resulting from fitting the flare intervals ($A-D$) are shown in Fig.~\ref{fig:highfit}.  The excess residuals between $7-8\keV$ are present in all the flare-intervals except for segment D, which has the highest flux, but shortest net exposure and fewest counts.

\begin{table*}
\caption{The best-fit model parameters for the flux resolved spectral modelling of \mrk335.  
The model components and model parameters are listed in Columns 1, and 2, respectively. 
Columns 3  to 8 list the parameter values during the flux-intervals labeled in Fig.~\ref{fig:lc}.
The inner ($R_{in}$) disc radius is assumed to extend to the innermost stable circular orbit (ISCO). 
The radius ($R_b$) at which the emissivity profile breaks from $q_{in}$ to $q_{out}$ is in units of gravitational radii 
($1\rg = GM/c^2$).  The reflection fraction ($\mathcal{R}$) is the ratio of the reflected flux over
the power law flux in the $0.1-100\keV$ band. 
Values that are linked between epochs appear only in column 3.  
The superscript $f$ identifies parameters that are fixed and the superscript $p$ indicates values that are pegged at a limit.
Fluxes are corrected for Galactic absorption and are reported in units of $\ergpscmps$.}
\centering
\scalebox{0.92}{
\begin{tabular}{ccccccccc}                
\hline
(1) & (2) & (3) & (4)  & (5) & (6) & (7) &  (8) \\
 Model Component &  Model Parameter  &  Average Low & Average High  & A & B & C & D \\
\hline
 Warm Absorber & $\nh$ ($\times10^{21}\pscm$) & $7.70\pm0.10$ &   \\
  ({\sc xstar})         &log$\xi$ [$\erg\cmps$] & $0.91^{+0.05}_{-0.04}$   \\
\hline
 Collisionally Ionised & kT ($\eV$) & $0.72^{+0.03}_{-0.02}$ &   \\
      Plasma ({\sc mekal})     &log $F_{0.5-10 keV}$ & $-12.99\pm0.02$ &   \\
\hline
 Power law & $\Gamma$ & $1.91^{+0.03}_{-0.02}$ & $2.15^{+0.04}_{-0.05}$  & $2.08^{+0.05}_{-0.06}$ & $2.07^{+0.04}_{-0.07}$ & $2.28^{+0.05}_{-0.07}$ & $2.18^{+0.08}_{-0.10}$ \\
  ({\sc cutoffpl})         &$E_{cut}$ ($\keV$) & $300^{f}$ &  \\
          &log $F_{0.1-100 keV}$ & $-11.97^{+0.05}_{-0.04}$ & $-11.37\pm0.04$ &  $-11.71\pm0.07$ &  $-11.41\pm0.06$ &  $-11.25\pm0.05$ & $-11.09\pm0.05$  \\
\hline
  Blurring  & $q_{in}$    & $6.38^{+0.20}_{-0.15}$   &  \\
  ({\sc kerrconv})         & $a$    & $>0.986$          &   \\
            & $R_{in}$ (ISCO)   & $1^{f}$         &                                \\
           & $R_{out}$   & $400^{f}$          &                                \\
           & $R_b$ ($\rg$)   & $6^{f}$ &  \\
           & $q_{out}$    & $3^{f}$          &   \\
           & $i$ ($\deg$)  & $52^{+4}_{-2}$  &                                      \\
\hline
  Reflection  & $\xi$ ($\erg\cmps$)   & $29\pm3$  & $38\pm9$   & $28^{+5}_{-4}$  & $36^{+17}_{-8}$  & $33^{+8}_{-6}$  & $50^{+12}_{-8}$   \\
({\sc reflionx})             & $A_{Fe}$ (Fe/solar)   & $3.74^{+0.26}_{-0.30}$          \\
             & $\mathcal{R}$ & $15.5^{+1.5}_{-1.9}$ & $4.8\pm{0.6}$  & $7.6^{+1.9}_{-1.7}$ & $6.5^{+1.4}_{-1.2}$ & $3.6\pm{0.6}$ & $3.2\pm{0.6}$\\
\hline
  Distant Reflector & $\xi$ ($\erg\cmps$)   & $1.0^{f}$  &          \\
 ({\sc reflionx})            & $A_{Fe}$ (Fe/solar)   & $1.0^{f}$  &         \\
     & $\Gamma$ & $1.9^{f}$ &  \\  
             &log $F_{0.5-10 keV}$ & $-12.13\pm0.03$    \\
\hline
  Average Flux &    $2-10\keV$ & $\approx2.79$  &  $\approx3.89$ &        \\
  ($\times10^{-12}\ergpscmps$) & $0.5-10\keV$    & $\approx3.56$   & $\approx5.96$   &          \\
\hline
              Fit Quality & $\Cdof$ & $851/606$     \\
\hline
\label{tab:fits}
\end{tabular}
}
\end{table*}

The same model simply scaled by a constant, fits the high- and low-flux MOS data equally well ($\Cdof=569/445$).  The background becomes significant above $\sim5\keV$, but residuals at $E\sim7.5\keV$ are apparent in the high-flux spectra that are not present in the low-flux spectra (Fig.~\ref{fig:meanfit}, lower panel).   An inverted Gaussian profile at $E=7.58^{+0.11}_{-0.34}\keV$ and with unconstrained width improves the residuals in the MOS spectrum.  An equivalent profile at $E=7.58^{+0.18}_{-0.34}\keV$ and $\sigma<300\eV$ improves the residuals in the pn spectrum.  The residuals in the MOS instruments appear statistically consistent with those detected in the pn.

The model is self-consistent across all flux intervals and in agreement with previous observations of the source.  Values of the distant reflector, warm absorber and ionised emitter are comparable to those measured in other works (e.g. Longinotti \et 2008, 2013, in prep; Grupe \et 2008; Gallo \et 2013, 2015) and do not appear to vary over the course of the observation.  Some narrow features are missed when applying the pn model to the RGS data.  The physical origin of the ionised emission features may be associated with some central starburst component, but it is difficult to determine if the emission arises in a collisional or photoionised plasma. On the other hand, the narrow emission features might also be originating from the distant narrow-line region.

The measurements of the blurring and accretion disc are in accord with the analyses of other data obtained at previous epochs and with different instruments (e.g. Grupe \et 2008, 2012; Gallo \et 2013, 2015; Parker \et 2014; Walton \et 2013).  In general, the X-ray emission from \mrk335 is consistent with originating from a compact corona, as noted by the steep inner emissivity profile ($q_{in}\approx6.4$) around a rapidly spinning black hole with a dimensionless spin parameter $a=cJ/GM^2 >0.986$, where $J$ is the angular momentum of a black hole of mass $M$.   The high spin measurement is consistent with all previous spin measurements for \mrk335\ and robust to the cautions raised by Bonson \& Gallo (2016).  The accretion disc, which extends to the innermost stable circular orbit (ISCO), is modestly ionised ($\xi\approx 30-50\erg\cmps$) and overabundant in iron by a factor of $\sim3.7$.  The reflection fraction ($\mathcal{R}$, the ratio between reflected and continuum flux between $0.1-100\keV$) is always greater than unity indicating the spectrum is dominated by the reflection component most likely because of light-bending effects (e.g. Miniutti \& Fabian 2004). 

\begin{figure*}
\begin{center}
\begin{minipage}{0.45\linewidth}
\scalebox{0.3}{\includegraphics[angle=90]{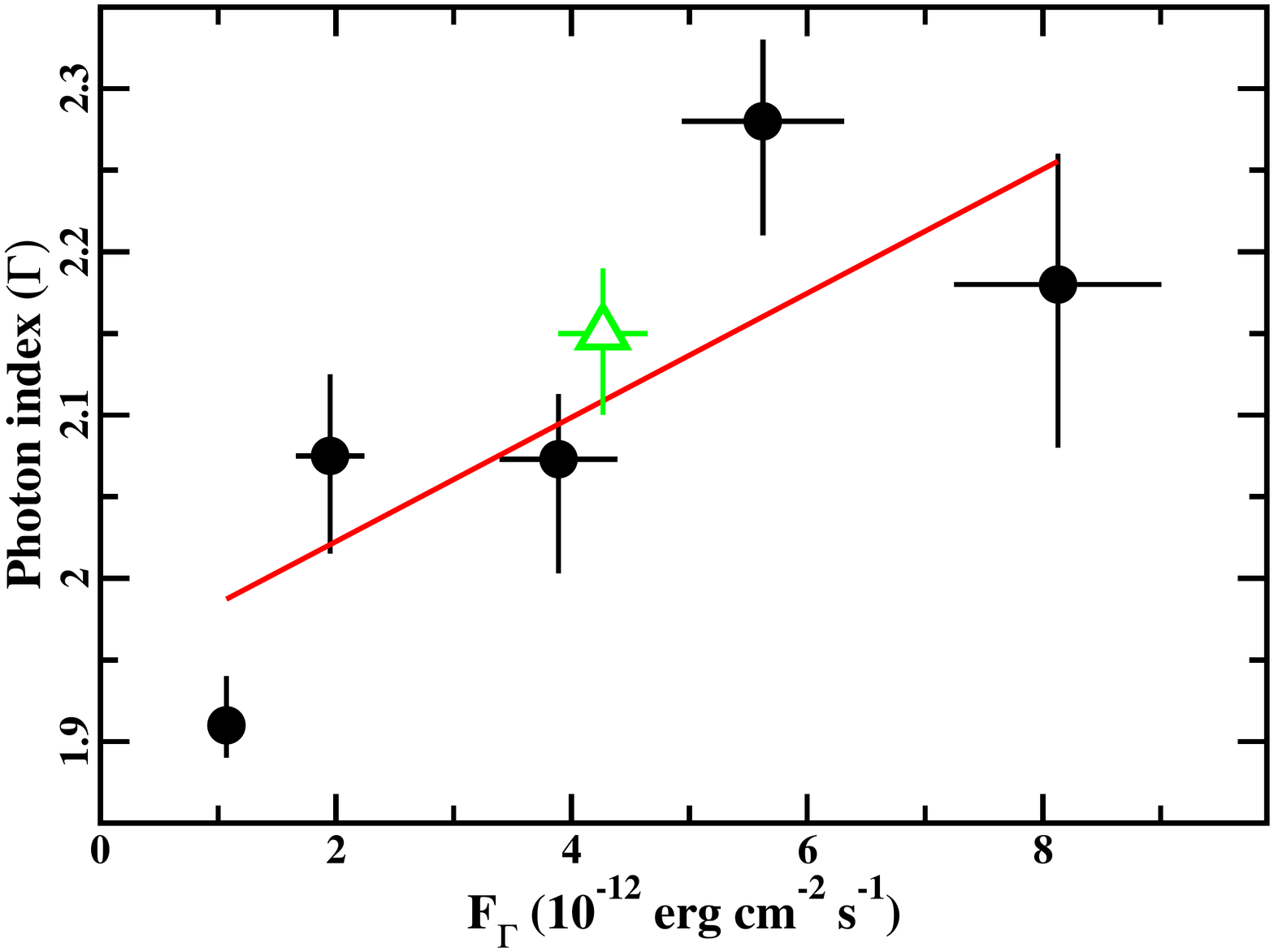}}
\end{minipage}  \hfill
\begin{minipage}{0.45\linewidth}
\scalebox{0.3}{\includegraphics[angle=90]{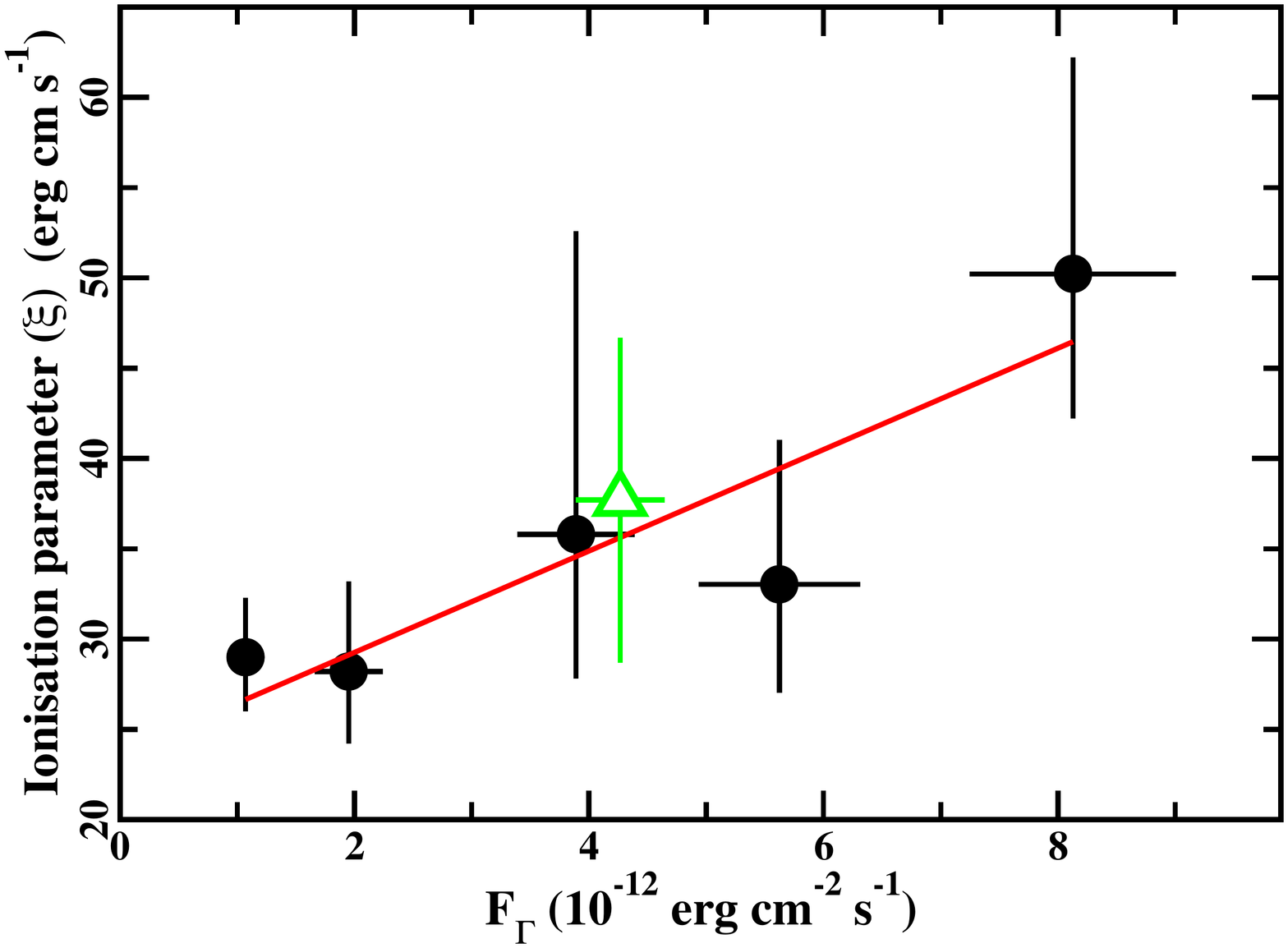}}
\end{minipage}  \hfill
\begin{minipage}{0.45\linewidth}
\scalebox{0.3}{\includegraphics[angle=90]{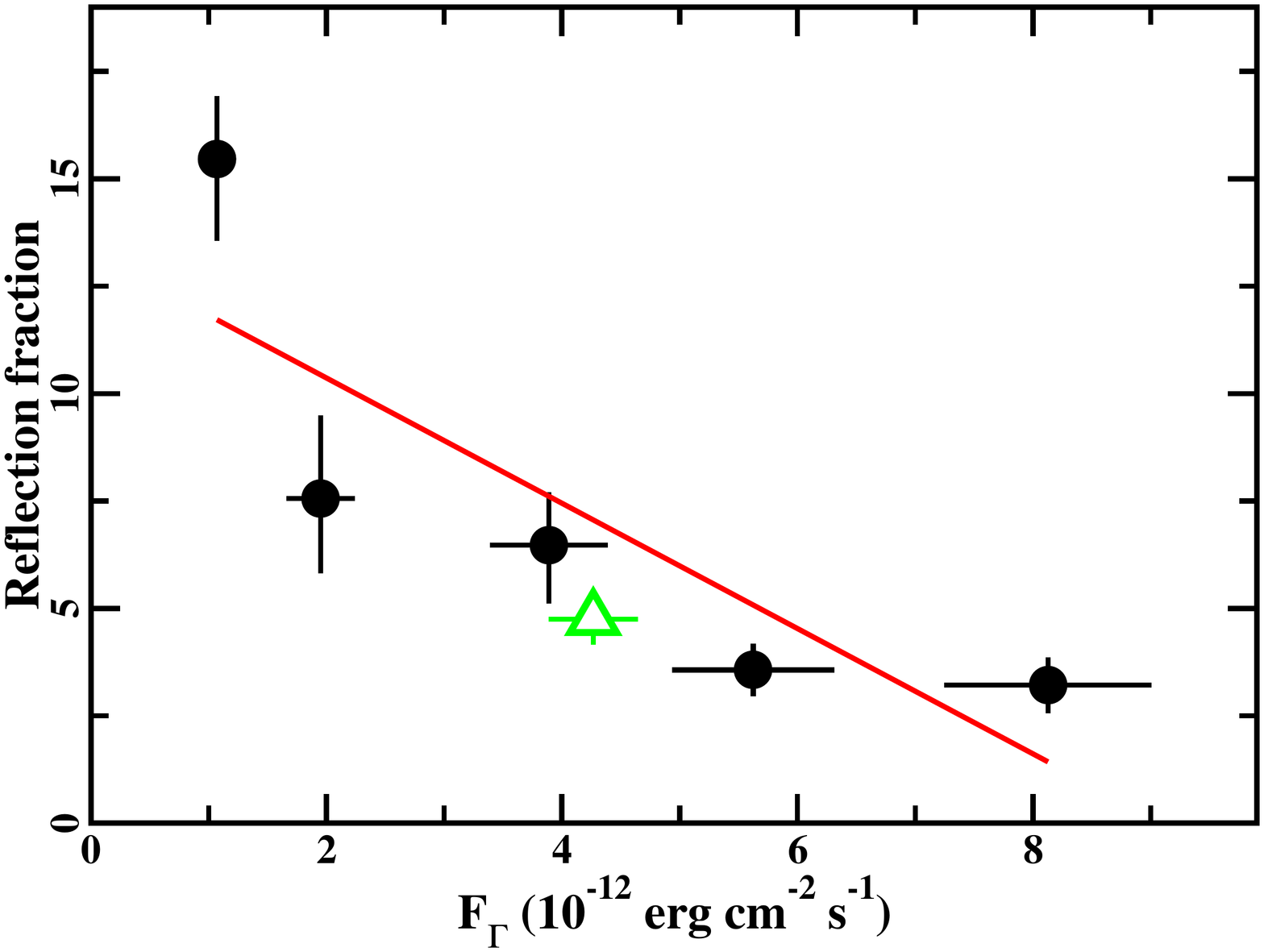}}
\end{minipage}  \hfill
\begin{minipage}{0.45\linewidth}
\scalebox{0.3}{\includegraphics[angle=90]{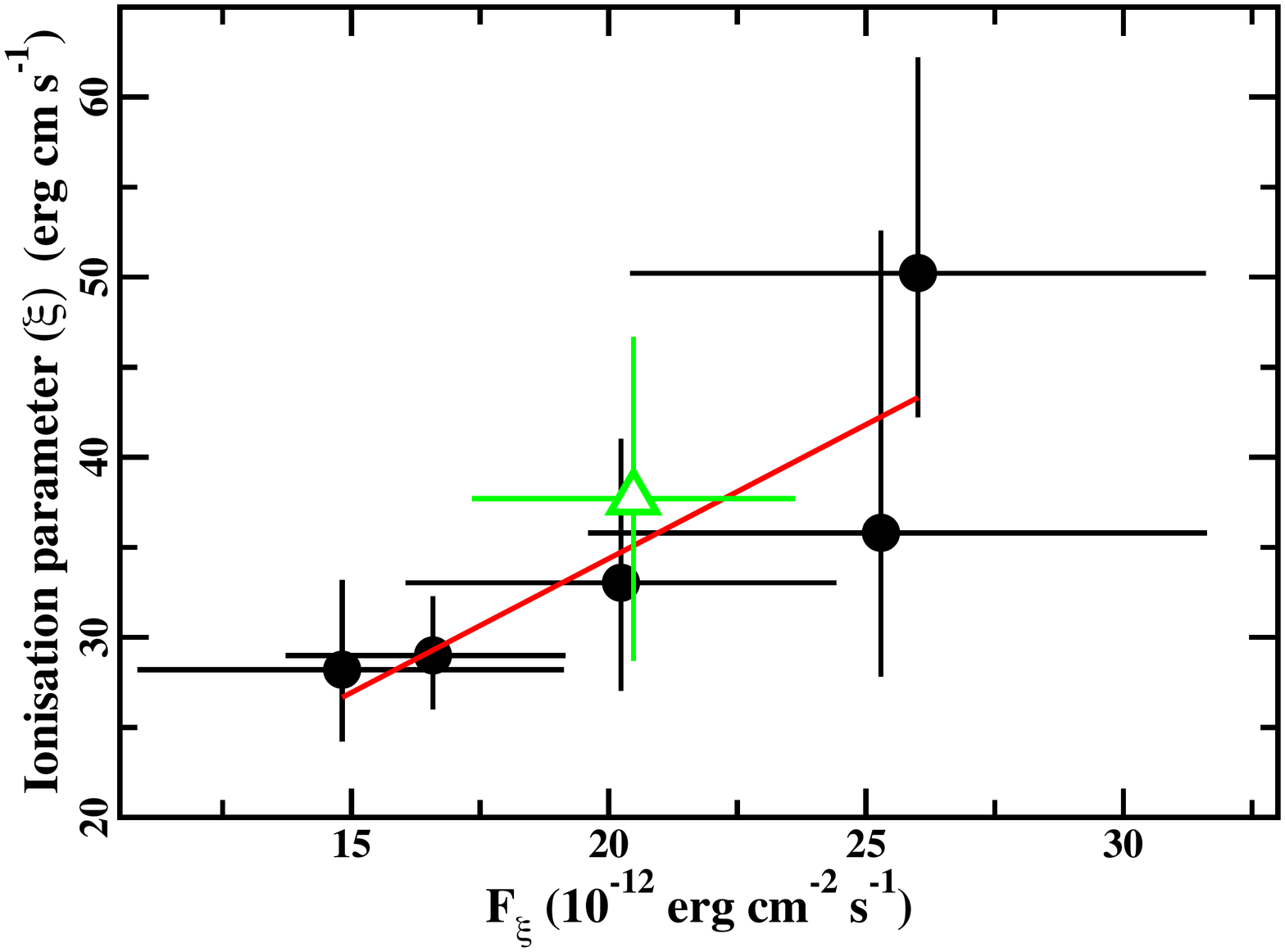}}
\end{minipage}
\end{center}
\caption{The observed correlations between the variable parameters in the blurred reflection model fitted simultaneously to the low-flux and high-flux intervals ($A-D$) (black, filled circles).  The solid, red curve corresponds to the best-fit line.  As the $0.1-100\keV$ power law flux increases:  the photon index steepens (upper left); the ionisation parameter increases (upper right); and the reflection fraction decreases (lower left).  The increase in ionisation parameter with increasing reflected flux is shown in the lower right. The measurements from the average flare (average high) spectrum is shown as a green open triangle.
}
\label{fig:corrs}
\end{figure*}

Correlations between the different parameters in the low and high-flux intervals (black, circles) are shown in Fig.~\ref{fig:corrs}.  A linear fit is applied to all the relations (solid, red curve) and in all cases the resulting correlation coefficient is high ($\approx \pm0.8$) indicating the trends are important.   The behaviour is as expected from previous studies of AGN variability and what is anticipated in the blurred reflection scenario.  

The power law photon index steepens with increasing power law flux (Fig.~\ref{fig:corrs}, top left) as is expected in accreting sources (e.g. Wang \et 2004).  The ionisation parameter (Fig.~\ref{fig:corrs}, top right) and reflected flux (Fig.~\ref{fig:corrs}, lower left) also increase as the power law flux rises during the flare.  This is expected if there is some positive correlations with the number of photons striking the disc and number of photons reflected into the observers line-of-sight.  The relation is corroborated in the lower right panel of Fig.~\ref{fig:corrs}, demonstrating the expected trend between ionisation parameter and reflected flux.  However, as the power law flux rises the reflection fraction ($\mathcal{R}$) diminishes suggesting that the relative contribution of the reflection spectrum becomes less important as the continuum brightens (Fig.~\ref{fig:corrs}).  This might be explained, though not exclusively, if the source is emitting anisotropically.

Keek \& Ballantyne (2016) investigated similar relations in multi-epoch data of \mrk335\ spanning $\sim14$ years.  There are slight differences in the models used and parameters investigated that make direct comparison difficult, but it is clear that similar trends exist in the long-term data.  Importantly, the Keek \& Ballantyne (2016) results support the notion that the corona geometry is changing and that corona expansion might be in the vertical direction (see also Ballantyne 2017). 

\subsection{Model-independent analysis}

The spectral variability is further examined by considering a flux-flux (e.g. Taylor \et 2003; Kammoun \et 2015) and principal component analysis (PCA) (e.g. Parker \et 2015).  

\begin{figure}
\rotatebox{0}
{\scalebox{0.37}{\includegraphics[trim= 0.9cm 0.8cm 1.5cm 2cm, clip=true]{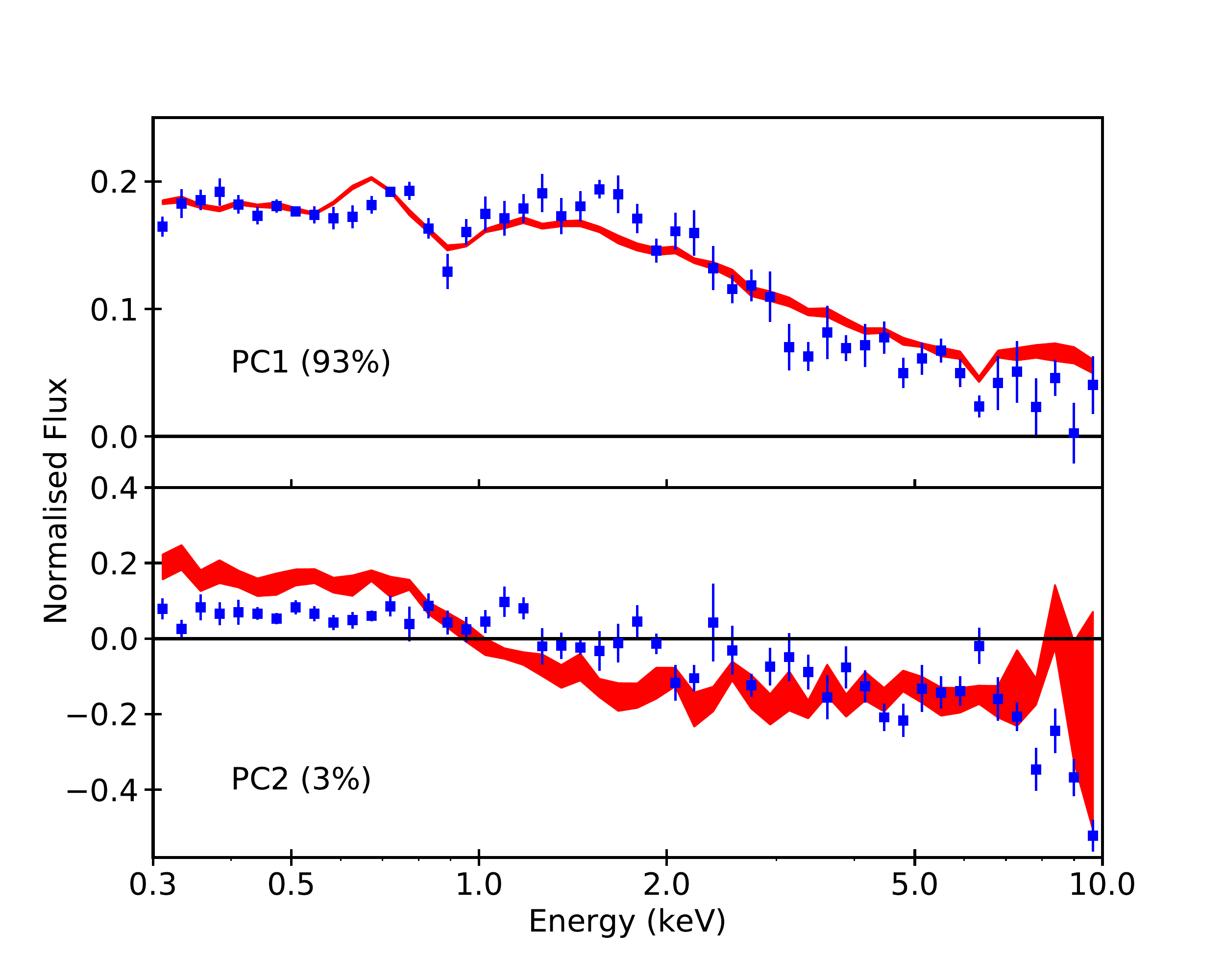}}}
\caption{
The principal component analysis reveals two significant eigenvectors (blue data points) (PC1 and PC2).  PC1 and PC2 account for 93 and 3 per cent of the variability, respectively.  The red curve shows the first two principal components that are measured assuming the best-fit model (Section~\ref{sect:spec}) in which the power law normalisation, photon index, and ionisation parameter vary in a correlated manner.  The width of the band indicates the uncertainty range in the model.
}
\label{fig:pca}
\end{figure}

The PCA is calculated for the 2015 observation using $5\ks$ intervals of the data.  When comparing the variability of each principal component on the log-eigenvalue diagram (LEV) as described in Parker \et (2015), the first two eigenvectors appear significant above the noise level.  The primary and secondary principal components (i.e. PC1 and PC2) account for approximately $93$ and $3$ per cent of the variability in the spectrum, respectively (Fig.~\ref{fig:pca}).  The shape of PC1 and PC2 as a function of energy can reveal the physical component in the source spectrum that is driving the variability.

As PC2 crosses zero normalisation (i.e. the normalisation is positive at low energies and negative at high energies), this indicates the high- and low-energy variations are anti-correlated.  Such behaviour is commonly associated with a pivoting power law (e.g. Parker \et 2015; Gallant \et 2018).  In contrast, PC1 shows variability that is correlated in all energy bands (i.e. all data points are positive), however the degree of variability is energy dependent.  Between $0.3-2\keV$, the variability is roughly comparable in each energy band, however as the energy increases above $\sim2\keV$, the variations diminish.

The spectral model above (Table~\ref{tab:fits}) is tested to determine if the scenario could account for the PCA.  Spectra are created using the best-fit model above and allowing the power law normalisation, photon index, and ionisation parameter to vary in a correlated manner.  The PCA is calculated from these simulated spectra that take into consideration the source brightness and background.

The PCA determined from the simulation are overplotted as a red band on the data in Fig.~\ref{fig:pca}.  The simulation reproduces the general shape of PC1 and PC2 relatively well.  Even sharp features in PC1 at energies of $\sim0.8-1\keV$ and $\sim6\keV$ that are likely associated with the warm absorber, emitter, or distant reflector are present and well-matched in the simulation.

\begin{figure}
\rotatebox{0}
{\scalebox{0.37}{\includegraphics[trim= 0.cm 0.cm 0cm 0cm, clip=true]{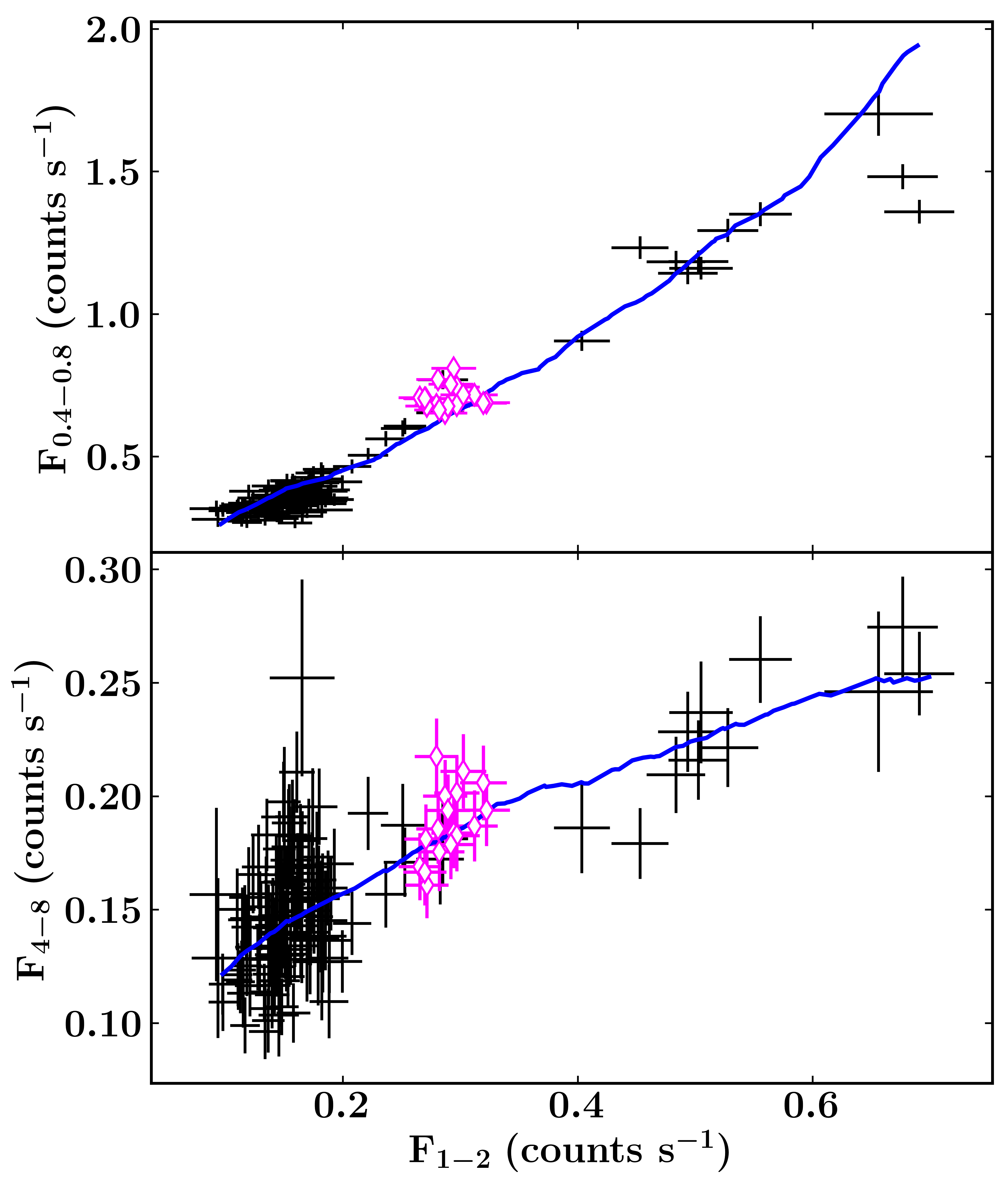}}}
\caption{
The flux-flux plots comparing the count rates in the soft band ($0.4-0.8\keV$, top panel) and hard band ($4-8\keV$, lower panel) to the $1-2\keV$ band.  The black data points are from the 2015 observation and the pink, open diamonds are data from the previous \xmm\ low-flux state observed in 2007.  The simulation used to model the PCA in Fig.~\ref{fig:pca} (i.e. power law normalisation, photon index, and ionisation parameter vary in a correlated manner) is adopted to generate the expected flux-flux relation that is overplotted on the data (solid blue curve).  
}
\label{fig:ffp}
\end{figure}

Flux-flux plots are created comparing a soft ($0.4-0.8\keV$) and hard ($4-8\keV$) band to the $1-2\keV$ band (Fig.~\ref{fig:ffp}).  Linear and power law models fitted to the flux-flux plots of \mrk335\ in the 2015 low state are of poor quality ($\redchi >2$).  The same set of simulated data used for the PCA were manipulated to create flux-flux plots that represent the scenario derived above.  These theoretical relations are overplotted on the 2015 data and show very good agreement (red curves in Fig.~\ref{fig:ffp}).  The data from the 2007 low-flux state, which did not exhibit marked variability during the observation, are also included in the flux-flux plot and appear completely consistent with the 2015 model.

The blurred reflection scenario describes well the flux-resolved spectra as well as the variability behaviour resolved with the PCA and flux-flux plots. The variability can be completely described by brightening of the power law component (i.e. the corona) that likely drives the correlated changes in the photon index and the accretion disc ionisation parameter.

\section{The low-state coronal geometry}
\label{sect:corona}
\begin{figure*}
\begin{center}
\begin{minipage}{0.45\linewidth}
\scalebox{0.45}{\includegraphics[trim= 0.cm 0cm 0cm 0cm, clip=true]{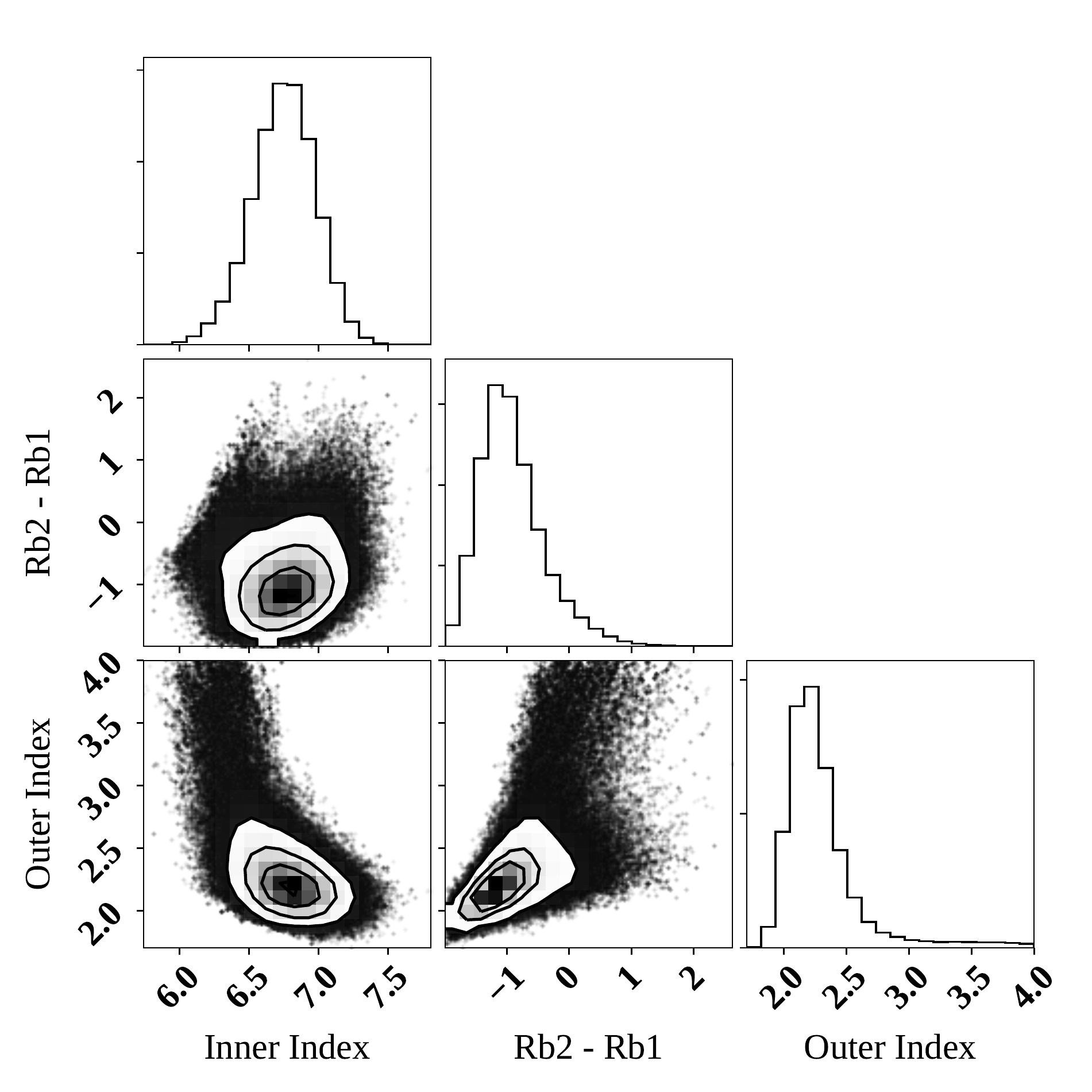}}
\end{minipage}  \hfill
\begin{minipage}{0.45\linewidth}
\scalebox{0.45}{\includegraphics[trim= 0.cm 0cm 0cm 0cm, clip=true]{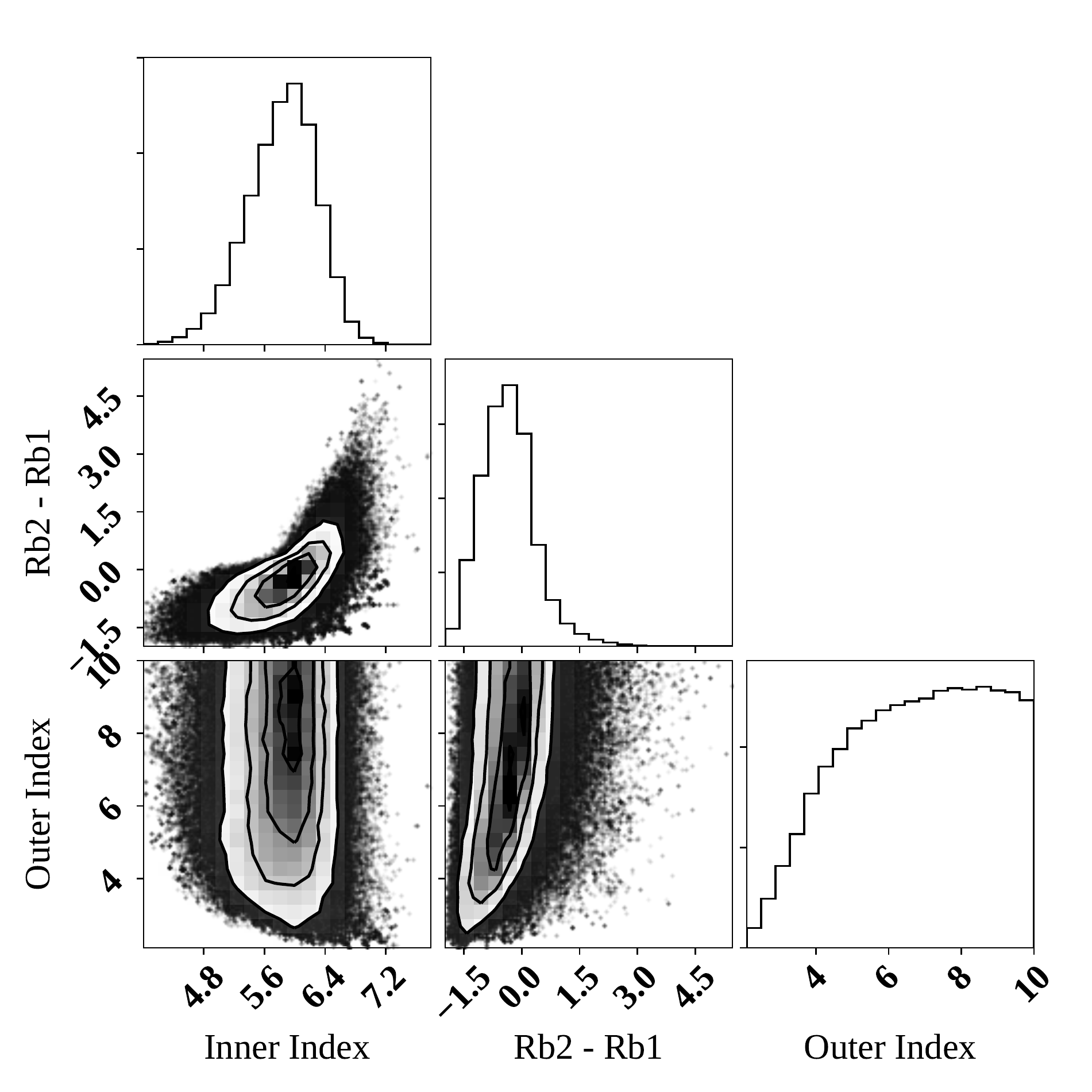}}
\end{minipage}  \hfill
\end{center}
\caption{The panels show the probability distribution and contours between the inner and outer emissivity index ($q_{in}$ and $q_{out}$, respectively) and inner and outer  break radius ($R_{b1}$ and $R_{b2}$, respectively) when fitting a twice-broken power law with a flat ($q=0$) profile in the middle region between the breaks.  In both the average low-flux data (left panel) and high-flux data (right panel), there is no indication of radial extension of the corona over the disc.
}
\label{fig:emis}
\end{figure*}
To probe the geometry of the corona, the emissivity profile of the accretion disc (that is the pattern of illumination of the disc by the coronal X-ray source) is measured. The emissivity profile of the disc is encoded in the profile of the relativistically broadened \feka\ line with the specific Doppler shift and redshift varying as a function of position on the disc, thus variations in reflected flux as a function of radius on the disc result in changes to the precise shape of the redshifted wing of the line.

We initially seek to measure the emissivity profile by fitting the profile of the \feka\ line as the sum of contributions from successive annuli on the disc, following the method of Wilkins \& Fabian (2011).   However the signal to noise in the detection of the line above the continuum during this single orbit \xmm\ observation in the low flux state is insufficient to constrain the emissivity of the disc at each radius.

Instead, we are guided by the suggested broken-power law emissivity profile in combination with the emissivity profile of the accretion disc measured in previous observations of \mrk335\ with \xmm\ in 2006, 2007 and 2009, \suzaku\ in 2006 and 2013, and \nustar\ in 2013 and 2014. The emissivity profile of the accretion disc was found to vary between a twice-broken power law, steeply falling in the inner regions, flattening to constant emissivity before breaking to the classically expected power law index of 3 over the outer disc, indicating a corona that is radially extended over the surface of the inner disc, during higher flux epochs to a once broken power law during the low flux epochs indicating illumination of the accretion disc by a compact X-ray source close to the black hole (Wilkins \& Gallo 2015; Wilkins \et 2015). The outer break radius indicates the radial extent of the corona over the disc (Wilkins \& Fabian 2012).

Instead of fitting a free function to the emissivity profile at each radius in the disc, we fit the profile of the \feka\ line with a single profile with a twice-broken power law emissivity profile. We use this model to test for evidence of any extension of the corona over the inner regions of the accretion disc. The power law index of the middle region of the profile is frozen at zero (the flattened portion of the profile produced by the extended portion of the corona), while the inner and outer power law indices and the two break radii are fit as free parameters to the observed line in the spectrum from $1-10\keV$. Having run an initial fit by minimising the C-statistic with respect to the free model parameters, we run Markov Chain Monte Carlo (MCMC) calculations from this starting point in parameter space to obtain an estimate of the probability distribution of each of the parameters of the emissivity profile. 
MCMC was performed using the Goodman-Weare algorithm implemented through {\sc xspec\_emcee},\footnote{Made available by Jeremy Sanders (http://github.com/jeremysanders/xspec\_emcee)} starting the walkers close to the best-fitting parameters found during the initial {\sc xspec} fit, perturbed by random increments with standard deviation from the covariance matrix found during the fit. The MCMC used 80 walkers, running for 10000 steps, burning the first 1000 steps of each chain.
We fit the model to the low flux state (before the flare) and the high flux state (during the flare) independently to understand the change in the corona as the flare is seen. Fig.~\ref{fig:emis} (centre and right panels) shows the probability distributions and contours between these parameters obtained from the MCMC analysis.

During both the low flux interval and the flare, the observed spectrum is consistent with there being no radial extension of the corona over the inner regions of the disc, hence the disc being illuminated by a central, compact source. This can be seen in the distribution in the difference between the break radii, showing the length of the flattened section (when $R_{b2} < R_{b1}$, the model produces simply a once broken power law breaking at $R_{b1}$ to the outer power law index). During the low flux interval, the difference in break radii is less than zero at the 94 per cent confidence level and less than 0.5 at the 98 per cent level. $R_{b2}$ is less than $5\rg$ at 96 per cent confidence. During the flare, the difference between the break radii is less than zero only at the 72 per cent confidence level, though less than 0.5 at the 90 per cent confidence level. The outer break radius is again less than $5\rg$ at 90 per cent confidence and less than $6\rg$ at 97 per cent confidence. It should be noted that from the twice-broken power law emissivity profile, it is not possible to measure radial extent of the corona below $5\rg$ since the inner steepening of the emissivity profile masks an outer break radius less than this, hence a measure of the break radius less than $5\rg$ is consistent with a compact corona that lies within this radius.

The most notable difference between the accretion disc emissivity profile between the low flux interval and the flare is the slope of the power law over the outer part of the accretion disc. During the low flux interval, the outer power law index is measured to be ($2.3\pm0.3$) revealing a flatter illumination profile over the outer disc than is expected typically from a compact X-ray source. Combined with the observation of a high reflection fraction, suggesting the bulk of the coronal emission is coming from close to the black hole such that the emission is focused onto the inner disc by strong light bending (e.g. Miniutti \et 2003; Suebswong \et 2006; Dauser \et 2013; Chainakun \& Young 2015), This flattening of the profile may be indicative of a slight vertical extension of the corona above the black hole as was seen during the 2006 observations of \mrk335\ with \suzaku\ (Wilkins \& Gallo 2015), though it is difficult to infer this from these data alone.

On the other hand, during the flare, the illumination of the outer disc falls off very steeply with radius, with the outer index greater than 3 at the 98 per cent confidence level and greater than 4 at the 90 per cent confidence level. This finding implies that little illumination reaches the outer parts of the accretion disc which would be expected in the case of mildly relativistic vertical motion of a centrally-collimated corona. Beaming of emission away from the disc results in few X-rays reaching the outer parts of the disc, hence producing the steep profile, while the inner disc is still strongly irradiated by X-rays that are bent down towards the black hole and inner disc in the strong gravitational field close to the black hole. Beaming of emission away from the disc also results in a low reflection fraction as the majority of the flux emitted by the corona is emitted upwards, being detected directly as continuum emission rather than being reprocessed by the disc.

\section{The onset of outflows}
\label{sect:outflow}
\begin{figure}
\rotatebox{0}
{\scalebox{0.55}{\includegraphics[trim= 0cm 0.cm 0cm 0cm, clip=true]{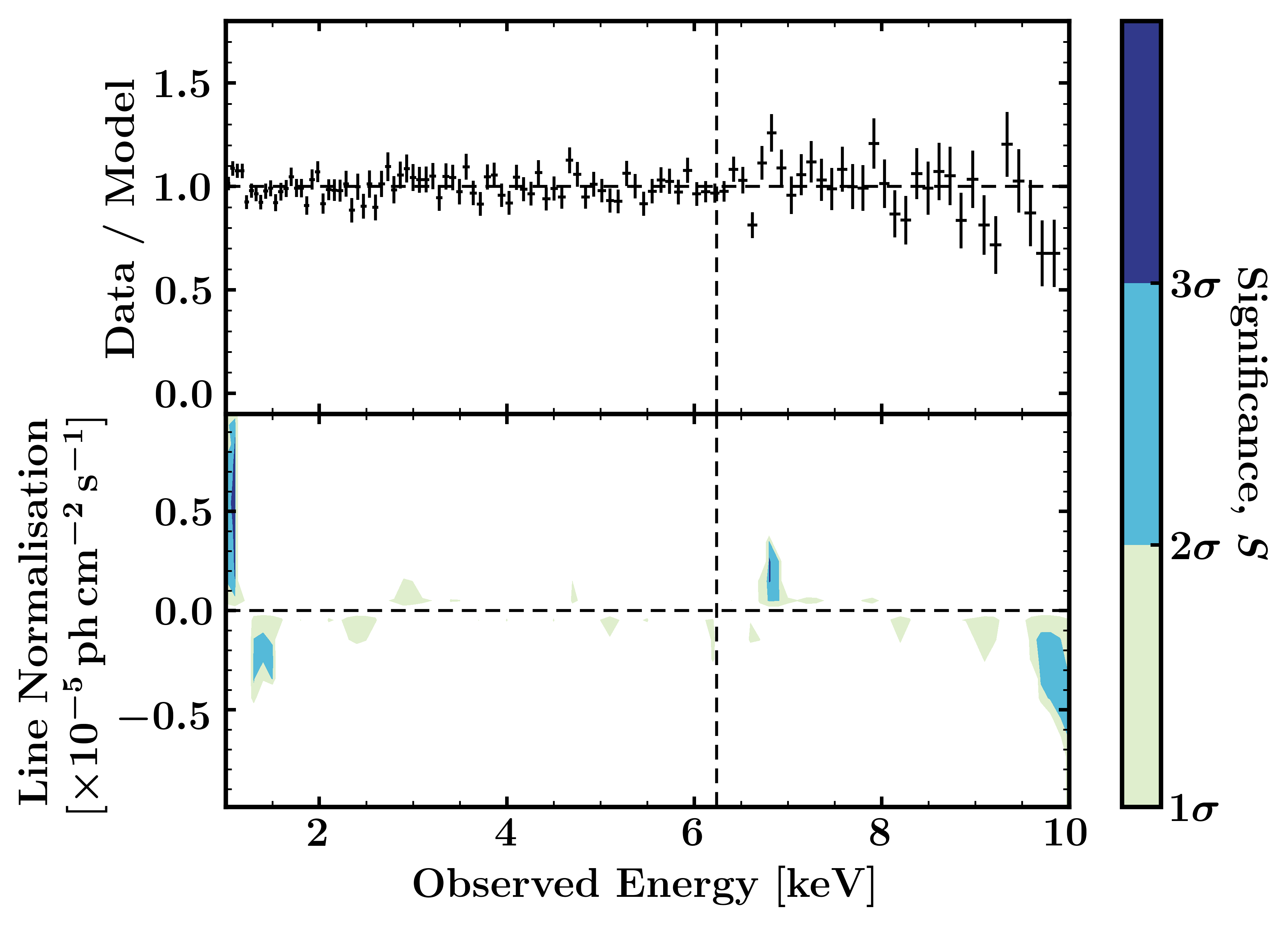}}}
{\scalebox{0.55}{\includegraphics[trim= 0.cm 0cm 0cm 0cm, clip=true]{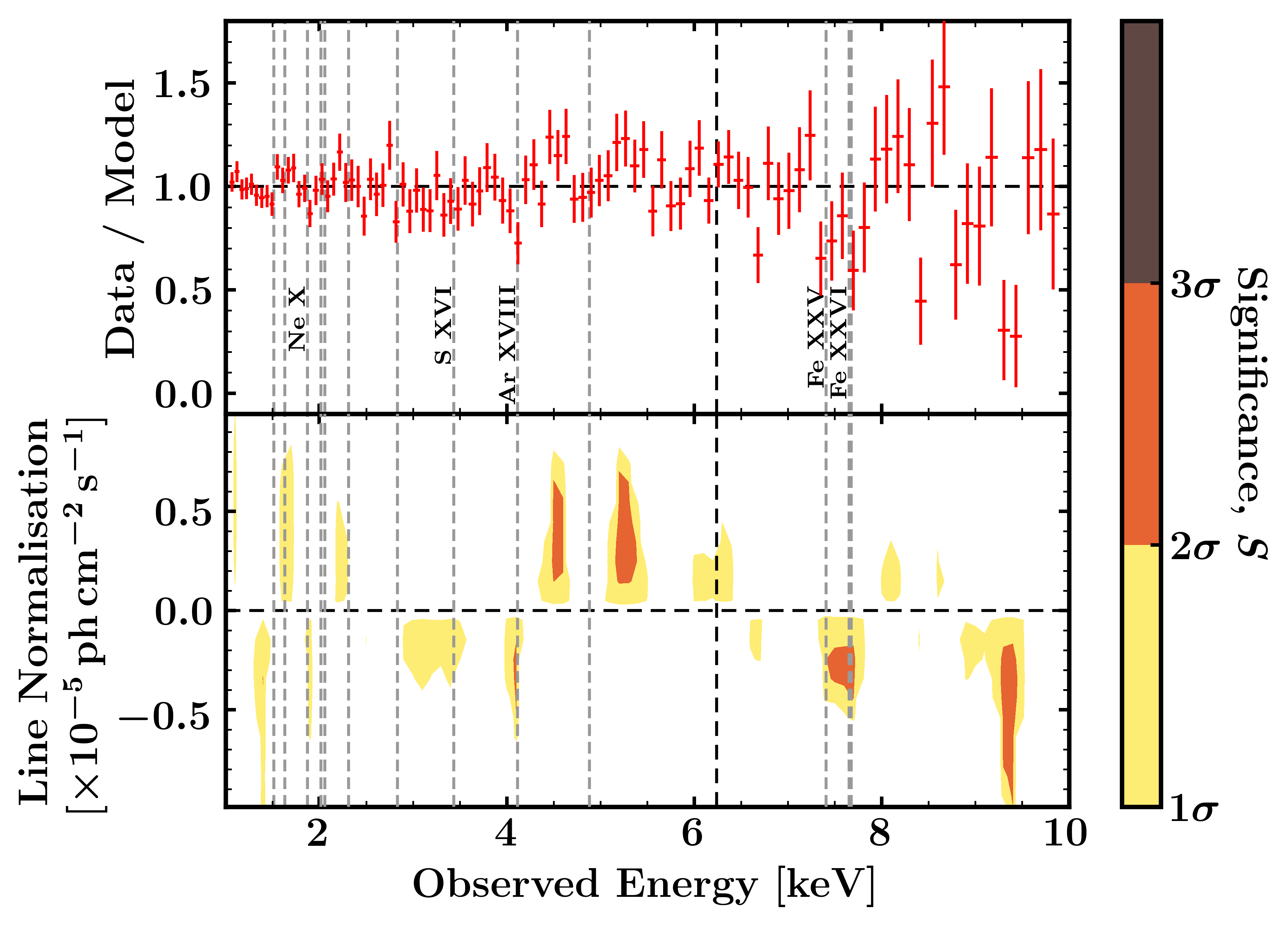}}}
\caption{
Each panel (top and bottom) includes two windows.  The first shows the ratio (data/model) plot in the $1-10$ keV band and the second shows the contours between line normalisation and energy of a Gaussian feature with width $10\eV$ produced from Monte Carlo simulations of the best fit model. The black vertical dashed line marks the observed Fe K$\alpha$ emission line energy of $6.4$ keV. Contour colours correspond to the indicated significance levels ($S$) in the colour bars.
The top panel depicts the search for features in the low-flux interval and the bottom panel shows the same for the high-flux (flaring) interval.
Included in the high-flux panels (bottom) are the line energies of strong features in a highly ionised gas ($\xi=10^4\erg\cmps$) that has been manually blueshifted from the rest-frame to best match the most significant contours.  The shift corresponds to a velocity of $0.12c$.
}
\label{fig:mc}
\end{figure}

Outflows have been previously reported for \mrk335\ (e.g. Longinotti \et 2008, 2013, in prep).  These outflows have intermediate velocities that are faster than those associated with warm absorbers ($\ls2000\kmps$), but slower than those defined as ultrafast ($\sim10^4\kmps$) outflows.
Highly ionised and ultrafast outflows have also been reported during a previous low X-ray flux state of \mrk335\ (e.g. see figures 3 and 4 of Parker \et 2014).  These different velocity winds could have a common origin if the lower velocity outflows are manifestations of the ultrafast outflows that collide with the interstellar medium (ISM) (e.g. King \& Pounds 2014). 

The inset in the lower panel of Fig.~\ref{fig:meanfit}, clearly shows negative residuals between $7-8\keV$ in the flaring spectrum.  Various continuum models are used to test the importance of the residuals.  The strength of the residuals do depend on the continuum model used, but they remain apparent in all cases.  The residuals could be attributed to highly ionised iron that is significantly blueshifted.

Before further discussion of implications, the statistical importance of these features is examined in a robust manner.  To determine the significance of features in the $1-10\keV$ band, a Monte Carlo simulation is carried out.  One-thousand spectra of the low- and high-flux states are simulated adopting the best-fit model (Table~\ref{tab:fits}), signal-to-noise, and optimal binning used to analyse the true data.  The high- and low-flux spectra are fit simultaneously over the entire $0.4-10\keV$ band and the residuals are examined between $1-10\keV$ at each flux interval.   Constraining a physical continuum model over the entire broadband is more meaningful than using a power law over a limited energy range to examine the residuals.  A Gaussian profile with a width of $10\eV$ is stepped by 0.1 keV steps through each spectrum separately and the improvement to the fit ($\Delta C$) is recorded, to determine the significance ($S$) of deviations at each energy.  The same technique is then applied to the real spectra resulting in the significance (contour) plots in Fig.~\ref{fig:mc}.  In the case of the real spectra, the model is linked between the flux states as in Section~\ref{sect:spec} (Table~\ref{tab:fits}), and a Gaussian profile is added and left free to vary in normalisation for the spectrum of interest while set to zero normalisation for the accompanying spectrum. 

Interestingly, between $1-10\keV$, the low-flux spectrum does not show any significant deviations beyond what might be expected from random statistical fluctuations (top panel, Fig.~\ref{fig:mc}).  However, the flare spectrum shows several deviations, both positive and negative, between the $2-3\sigma$ level (lower panel, Fig.~\ref{fig:mc}).  Above $\sim9\keV$, residuals are regarded less important as the background level becomes noteworthy (Fig.~\ref{fig:meanfit}).  

In Fig.~\ref{fig:mc}, there are residuals between $2-3\sigma$ seen in absorption in the $7-8\keV$ band.  Features in this energy band are often attributed to outflowing He- and H-like \feka.  The addition of two narrow ($\sigma=1\eV$) Gaussian absorption features at fixed energies of $6.7$ and $6.97\keV$ fit the residuals well with a common blueshift velocity of $v=0.12^{+0.08}_{-0.04}c$. 

A highly ionised gas producing strong He- and H-like \feka\ would also generate other spectral features from species at lower atomic numbers.  An {\sc xstar} grid with variable column density, ionisation parameter, and iron abundance is fitted to the high-flux spectrum to physically model the residuals.  The inclusion of the grid improves the residuals, but is not well-constrained with the current data quality.  The best fit parameters of the outflowing absorber are $\nh\approx10^{23}\pscm$, $\xi\approx10^4\erg\cmps$, $A_{Fe}=2$ solar, and $v=0.12\pm0.02c$. The stronger features produced by such a plasma, blueshifted by a velocity of $v\sim0.12c$ could reproduce some of the contours as indicated in Fig.~\ref{fig:mc}.

Comparable features were reported in the 2013 \nustar\ observation by Parker \et (2014).  Unfortunately, in the \suzaku\ low-state observations that most closely resembles these current data in flux (Gallo \et 2015), the spectral region was background dominated by a Ni feature at $7.5\keV$. 

Features of comparable $\sim2\sigma$ significance are seen in emission at lower energies of \feka.  Narrow Gaussian profiles can be fitted at $4.66^{+0.08}_{-0.09}$ and $5.38^{+0.11}_{-0.09}\keV$.  The measured values correspond to rest-frame energies of Ti~K$\alpha$ and Cr~K$\alpha$, respectively.    However, the features are much stronger than would be expected in a plasma of cosmic abundances and could just be resulting from an imperfect fit to the relativistic broad line.

\section{Discussion} 
The 2015 \xmm\  observation of  \mrk335\ caught the NLS1 in its lowest X-ray flux state since the AGN became X-ray weak in 2007 (Grupe \et 2007, 2008).  This extended X-ray weak state has been marked with continuous flickering in the X-ray and UV light curves, and occasional high amplitude X-ray flares in which the NLS1 can become $\sim50\times$ brighter (e.g. Grupe \et 2013; Wilkins \et 2015; Gallo \et 2018).  

Since the onset of the monitoring in this X-ray weak state, there have been suggestions (Gallo \et 2013) that the X-ray flaring in \mrk335\ could be attributed to the base of an aborted jet (e.g. Ghisellini \et 2003).  The jet-like corona may have even prevailed when \mrk335 was in the stable, X-ray bright state prior to 2007.  During that epoch the spectra were dominated by the power law component and exhibited low reflection fractions ($\mathcal{R}<1$) (e.g. Gallo \et 2015; Wilkins \& Gallo 2015), which could be attributed to a corona moving at high velocity away from the disc.

The interpretation relies on the determination of the emissivity profile and simultaneous measurement of the reflection fraction when the source flares.  In some NLS1s, the brightness is associated with the radial expansion of the corona over the disc.  For instance, in \1h07, when the source was bright, the corona extended radially outwards, whereas when dim the source was confined to a small region close to the black hole  (Wilkins \et 2014).

However, in previous observations of \mrk335, the corona presumably extends vertically during flaring events.  At the same time, the reflection fraction ($\mathcal{R}$) is seen to decrease even to values less than unity.  This would indicate that the corona is  illuminating anisotropically, preferentially away from the accretion disc. This might be expected from a corona that is moving at high velocities away from the disc and whose illumination is beamed (e.g. Beloborodov 1999; Gonzalez \et 2017b).  This behaviour leaves signatures in the reverberation lag-energy spectrum (Wilkins \et 2016) that have been observed in \izw1\ (Wilkins \et 2017), another NLS1 that shares a similar interpretation for its X-ray behaviour (e.g. Gallo \et 2007; Gallo 2018).  Such timing signatures have not been reported in \mrk335\ since recent flaring observations were obtained with \suzaku\ and \nustar, which lack the timing resolution to carry out such work effectively.

During the 2015 observation, \mrk335\ remained quiescent for the first $\sim120\ks$ before flaring in the last $\sim20\ks$.  The end of the observation truncated the flare as it was still  brightening (Fig.~\ref{fig:lc}).  Though the complete flare was not recorded, the behaviour of \mrk335\ during this time was completely consistent with previous observed flares in the AGN.  The source remained compact as it brightened, but the emissivity profile of the outer disc steepened indicating less illumination was reaching the disc at large distances (Section~\ref{sect:corona}).  The power law spectrum softened (dropping in temperature or opacity) indicating the corona could be extending vertically since no radial extent was measured.   Simultaneously, the reflection fraction dropped significantly from $\sim15$ to $\sim3$ in about $20\ks$ (Section~\ref{sect:spec}). 

Values of $\mathcal{R}<1$ that would indicate beamed emission are never measured during the observation.  However, in conjunction with previous analyses, the rapid decline in $\mathcal{R}$ is still supportive of the aborted jet scenario.  During the flare the dynamic corona in \mrk335\ may be collimating and moving away from the disc.  Alternatively, the corona could also have a blended configuration as suggested for \izw1\ (e.g. Gallo \et 2007; Wilkins \et 2017), but it would still need to be rather compact.

\begin{figure}
\rotatebox{0}
{\scalebox{0.55}{\includegraphics[trim= 0cm 0cm 0cm 0cm, clip=true]{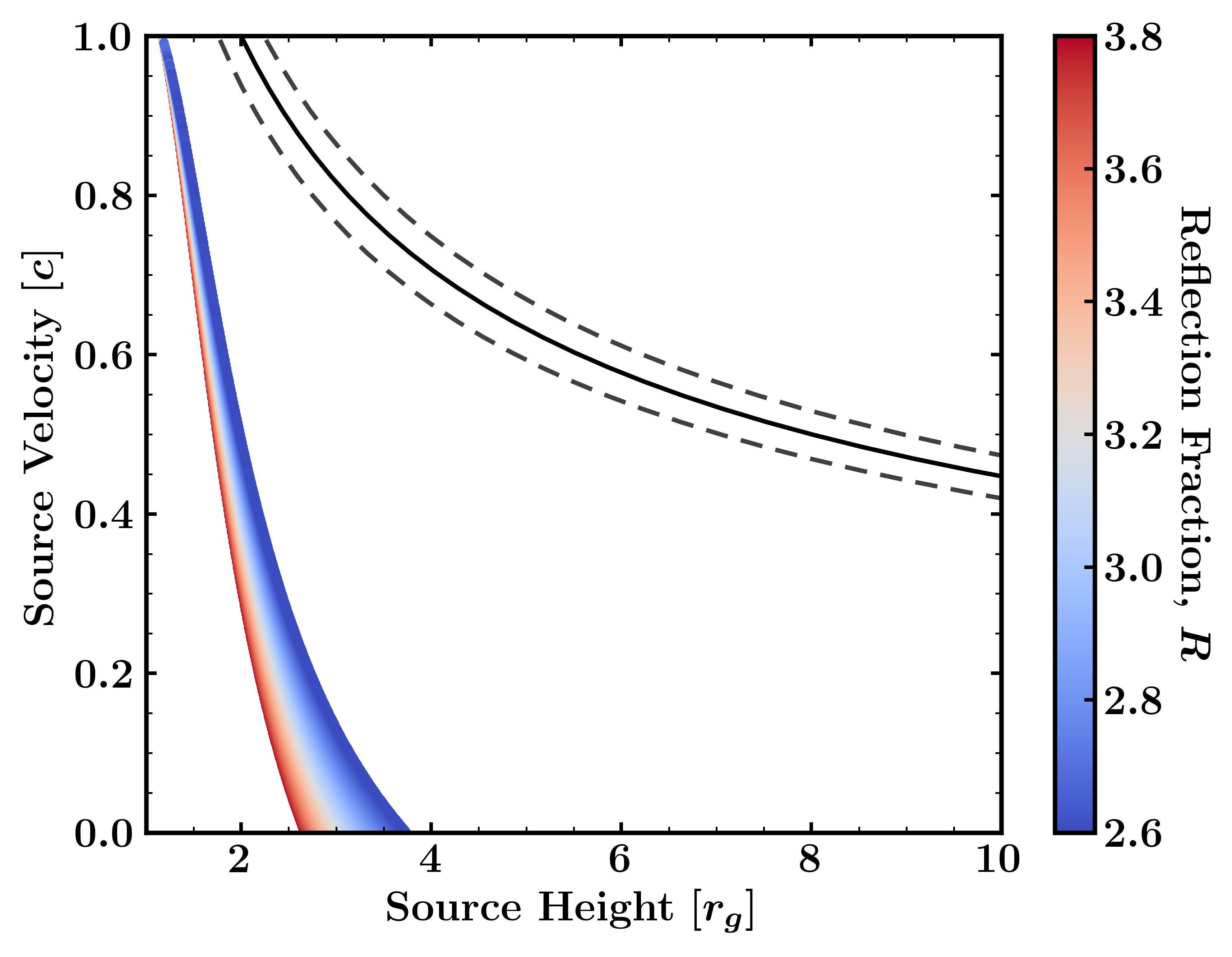}}}
\caption{
Based on the lowest measured reflection fraction ($\mathcal{R}\approx3$), the velocity
of the ejecta in the collimated corona as a function of height can be estimated (colour band). Compared to the escape
velocity assuming the estimated black hole mass for \mrk335\ (solid and dashed curves) the ejecta will not escape the system.
}
\label{fig:rzb}
\end{figure}

While it is not certain if the reflection fraction continues to fall after the observation ends, the value of $\mathcal{R}\approx3$ can be used to estimate the velocity of the corona as a function of height ($h$) above the black hole (equation 15 of Gonzalez et al. 2017b) (Fig.~\ref{fig:rzb}). Under the assumptions of a maximally spinning Kerr black hole and a compact source geometry aligned with the spin axis of the black hole (i.e. ``lamp-post'' geometry), we may use the measured value of $\mathcal{R}$ at the peak of the flaring event to evaluate the source velocity (i.e. corona outflow velocity) at different source heights above the black hole. As the final measured value of $\mathcal{R}$ is still relatively high, we find the source remains close to the black hole ($h\ls4\rg$) for all possible velocities. 

The escape velocity from the system has also been calculated using the black hole mass estimate for \mrk335\  of $2.5\pm0.3 \times 10^7 M_\odot$ (Grier \et 2012; similar mass as measured by Du \et 2014). For all combinations of velocity and height, the source never exceeds the escape velocity of the system (Fig.~\ref{fig:rzb}) so the material will likely fall back toward the accretion disc. Note, that despite the assumption of a compact source, Gonzalez \et (2017b) do not find the results for a collimated source to be significantly different.

The conjectured collimated, outflowing corona, may also be accompanied by a wind-like outflow (Section~\ref{sect:outflow}).   The investigation of ultrafast outflows often relies on the presences of absorption features at energies above $\sim7\keV$.  During the quiescent phase of the observation ($\ls120\ks$) no such features were present in the spectrum.  However, during the flaring event, a number of features were detected between the $2-3\sigma$ level.  Residuals were seen in emission and absorption, but an absorption feature at $\sim7.5\keV$ in the observed frame could be be attributed to highly ionised iron in outflow (Fig.~\ref{fig:mc}). Similar residuals at $\sim7.5\keV$ were also seen in the 2013 \nustar\ observation of \mrk335\ (e.g. see figures 3 and 4 of Parker \et 2014) . Comparison of the absorption features with a highly ionised plasma ($\xi=10^4\erg\cmps$) describe the data reasonably well with a single velocity shift of $v\sim0.12c$.  The feature at $7.5\keV$ is physically described as arising from both He- and H-like iron.  Other residuals in Fig.~\ref{fig:mc} might also be attributed with Ar~\textsc{xviii}, S~\textsc{vi}, and Ne~\textsc{x}.   

Alternatively, a disc origin for the absorption features seen in some AGN was proposed by Gallo \& Fabian (2011, 2013).  Similar spectral features could be reproduced if the ionising material were corotating with the disc rather than outflowing.  However, in such a scenario, the absorption features would become stronger with increasing reflection fraction, which is in contrast to what is observed in \mrk335.

The variability of the possible ultrafast wind in \mrk335\ is different than that described for the possible wind in another NLS1, \iras13\ (Parker \et 2017; Pinto \et 2018; Jiang \et 2018).  In \iras13, which is likely accreting near its Eddington limit, the wind features were seen when the source was dim and not when is was bright.  Parameters other than radiation pressure might be driving the wind (e.g. King \& Pounds 2003).  The interpretation was that the wind is perhaps always present, but over-ionised when \iras13\ was bright and therefore no wind features were observed.  This is opposite of what is observed in \mrk335, which is a sub-Eddington source (i.e. $L_{bol}/L_{Edd} < 1$) in the low-flux state (e.g. Keek \& Ballantyne 2016).  Here, like in X-ray binaries, the potential ultrafast outflow features are seen during the bright, flaring state and not when the source is dim.   It could be that the wind in \mrk335, is more massive and simply requires more energy to be moved and that the radiation pressure is not sufficient to launch the wind during the quiescent state and can only do so when the flare occurs.  

On the other hand, the absorption features were not reported in the pre-2007 observations when \mrk335\ was $\sim10$-times more luminous (Crummy \et 2006; Longinotti \et 2007, 2013; Larsson \et 2007; Gallo \et 2015; Wilkins \& Gallo 2015) despite evidence that the high-flux state maybe a more extreme example of the flare states.  It could be that in the bright state (pre-2007 or prolonged flaring events since 2007), \mrk335\ behaves like \iras13 and over-ionises the purported wind.

If the flare originates in the compact corona as proposed, than the wind must be launched from nearby.  The absorption features are reported in the average flare spectrum and they are even less significant in the short $\sim5\ks$ flare intervals.  This would imply that the features appear on time scales of $<10\ks$ from the onset of the flare, which would correspond to a light-travel distance of $<80\rg$ from the compact corona, assuming a mass of $2.5\times10^7\Msun$ (Grier \et 2012) for the black hole in \mrk335.  The escape velocity at $80\rg$ is $\sim0.15c$ so such a wind could be ejecting energy out of the AGN environment.  

In the scenario described by King \& Pounds (2014), if this wind were to collide with the ISM, shocks could generate the fast warm absorber velocities reported by Longinotti \et (2013).  It is worth noting that the star formation rate relative to AGN luminosity in \mrk335\ is the lowest seen in NLS1s (Sani \et 2010).  This could be indication that the star formation in \mrk335\ has been suppressed by past activity that was similar to what we are witnessing currently.

\section{Conclusions } 
 
 A triggered $140\ks$ \xmm\ observation of the NLS1 \mrk335\ finds the AGN in a relatively quiescent phase in the first $120\ks$ before flaring.  The data are compared with the blurred reflection model.  The spectrum is consistent with the blurred reflection scenario in which the corona is compact around a Kerr black hole and light bending is important.  The variability analysis (flux-flux plots and PCA) as well as flux-resolved spectroscopy demonstrates that the variability can be attributed to brightening and softening of the power law component (i.e. the corona) and increased ionisation of the inner accretion disc.  Of particular interest is that during the flare, the primary continuum is consistent with originating from a corona that might be extending vertically and beaming away from the disc, and that features appearing in the spectrum could arise in an accretion disc wind outflowing at $v\sim0.12c$.  If confirmed, this is the first example of a radio-quiet AGN exhibiting behaviour consistent with diffuse and collimated outflow.  Understanding such objects is necessary to determine the origin of the launching mechanism of winds and jets.
 

\section*{Acknowledgments}
The \xmm\ project is an ESA Science Mission with instruments
and contributions directly funded by ESA Member States and the
USA (NASA).  We are grateful to the \xmm\ observing team for preparing and
activating the ToO.  Many thanks to the referee for a thoughtful review and helpful comments.




\bsp
\label{lastpage}
\end{document}